%% file: MRFM_MagneticNoise.tex
\documentclass[11pt,oneside]{amsart}
\usepackage{MRFM_AMSLaTeX_macros}	
\usepackage{cite}					
\usepackage{xspace}					
\numberwithin{equation}{section} 	

\input{epsf}

\newcommand{\thirtyoneP}{\ensuremath{\sp{31}\text{P}}\xspace}
\newcommand{\oneninenineHg}{\ensuremath{\sp{199}\text{Hg}}\xspace}

\newcommand{\Stieltjies}{%
   \mbox{Stielt
   \JASkern{-0.04em}j
   \JASkern{ 0.05em}i
   \JASkern{ 0.02em}e
   \JASkern{ 0.02em}s}
   \xspace}
\newcommand{\JASkern}[1]{\hspace{#1}}  

\begin{document}


\title[Fluctuation, Dissipation, and Entanglement]
{Fluctuation, Dissipation,
and Entanglement:\\
the Classical and Quantum Theory 
of Thermal Magnetic Noise} 
\author{
J.\ A.\ Sidles\\
J.\ L.\ Garbini\\
W.\ M.\ Dougherty\\
S.\ H.\ Chao}
\address{
		University of Washington \\
		Department of Orthop{\ae}dics \\ 
		Seattle, Washington 98195} 
\email{sidles@u.washington.edu} 
\thanks{Supported by the NIH Biomedical Research Technology Program 
(BRTP), the NSF Major Research Instrumentation Program (MRI), the 
U.S.\ Army Research Office (ARO), and the University of Washington 
Department of Orthop{\ae}dics.}

\date{\today}

\begin{abstract}
A general theory of thermal magnetic fluctuations near the surface of
conductive and/or magnetically permeable slabs is developed; such
fluctuations are the magnetic analog of Johnson voltage noise. 
Starting with the fluctuation-dissipation theorem and Maxwell's
equations, a closed-form expression for the magnetic noise spectral
density is derived.  Quantum decoherence, as induced by thermal
magnetic noise, is analyzed via the independent oscillator heat bath
model of Ford, Lewis, and O'Connell.  The resulting quantum Langevin
equations yield closed-form expressions for the spin relaxation times
$T_{1}$, $T_{2}$, and $T_{1\rho}$.  For realistic experiments in
atomic physics, quantum computing, and magnetic resonance force
microcopy (MRFM), the predicted relaxation rates are rapid enough that
substantial experimental care must be taken to minimize them.  At zero
temperature, the quantum entanglement between a spin state and a
thermal reservoir is computed.  The same Hamiltonian matrix elements
that govern fluctuation and dissipation are shown to also govern
entanglement and renormalization, and a specific example of a
fluctuation-dissipation-entanglement theorem is constructed.  We
postulate that this theorem is independent of the detailed structure
of thermal reservoirs, and therefore expresses a general thermodynamic
principle.
\end{abstract}

\maketitle
\newpage
\tableofcontents

\include{MRFM_Magnetic_Part_01}

\appendix
\include{MRFM_Magnetic_Part_02}

\input{MRFM_Bibliography}
\newpage
\begin{figure}
\begin{center}
	 \epsfxsize=2.5in \epsfbox{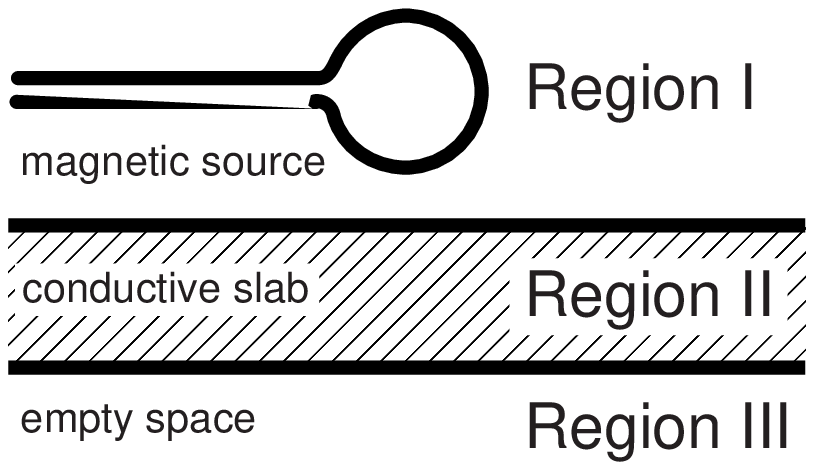}
	 
	 \caption{Conductive slab geometry.  For a realistic rendition of 
	 B-field and E-field geometry, see Fig.~\ref{fig:SkinCoil}.  This 
	 figure defines Regions I, II, and III as referenced in 
	 \eqref{eq:LinearEquations}.}
	 
\label{fig:Regions}
\end{center}
\end{figure}

\begin{figure}
\begin{center}
\epsfxsize=4.5in 
\epsfbox{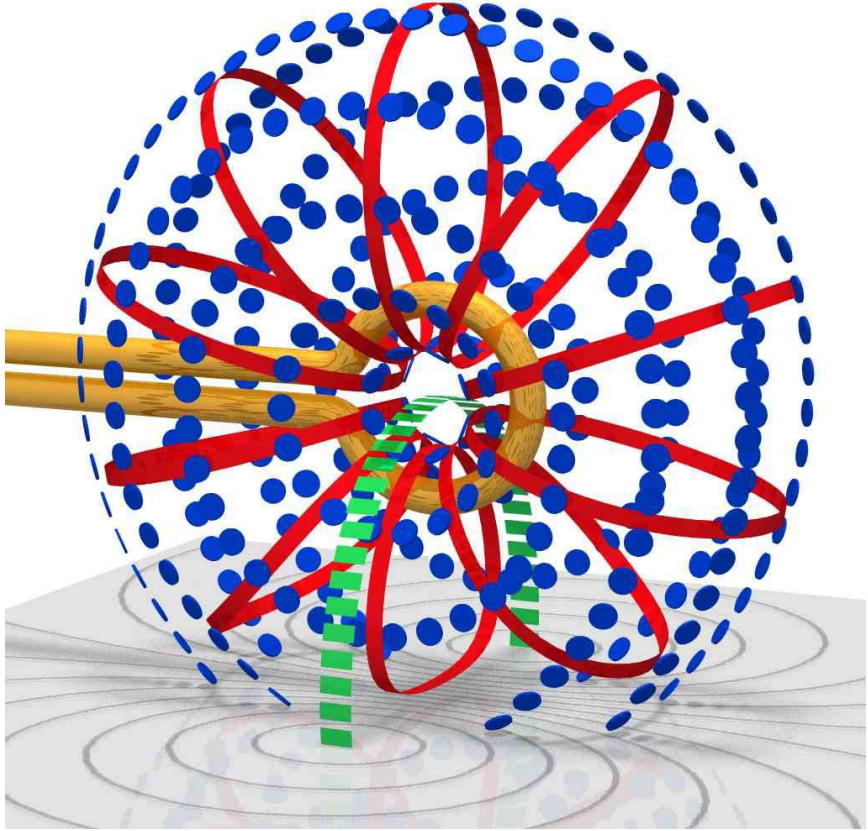} 

\caption{The B-field, E-field, and intra-slab currents---drawn to 
accurate scale---created by external excitation of a finite-sized 
current loop, as specified by (\ref{eq:A}--\ref{eq:finiteFields}).  
The field geometry has a ready physical interpretation: the coil 
excitation current generates a B-field (solid ribbons), 
whose time-rate-of change induces a toroidal E-field (circumferential 
dashed ribbons), that in turn induces currents within the 
slab (streamlines on the slab surface) which create a 
return flux loop (broad dashed ribbon) linking the induced 
slab currents back to the coil.  The return flux induces a 
phase-lagged coil voltage equivalent to that of a resistive impedance, 
thus ensuring that the energy externally supplied to excite the coil 
balances the energy dissipated in the slab.
}

\label{fig:SkinCoil}
\end{center}
\end{figure}

%
\begin{figure}
\begin{center}
	\epsfxsize=4.5in \epsfbox{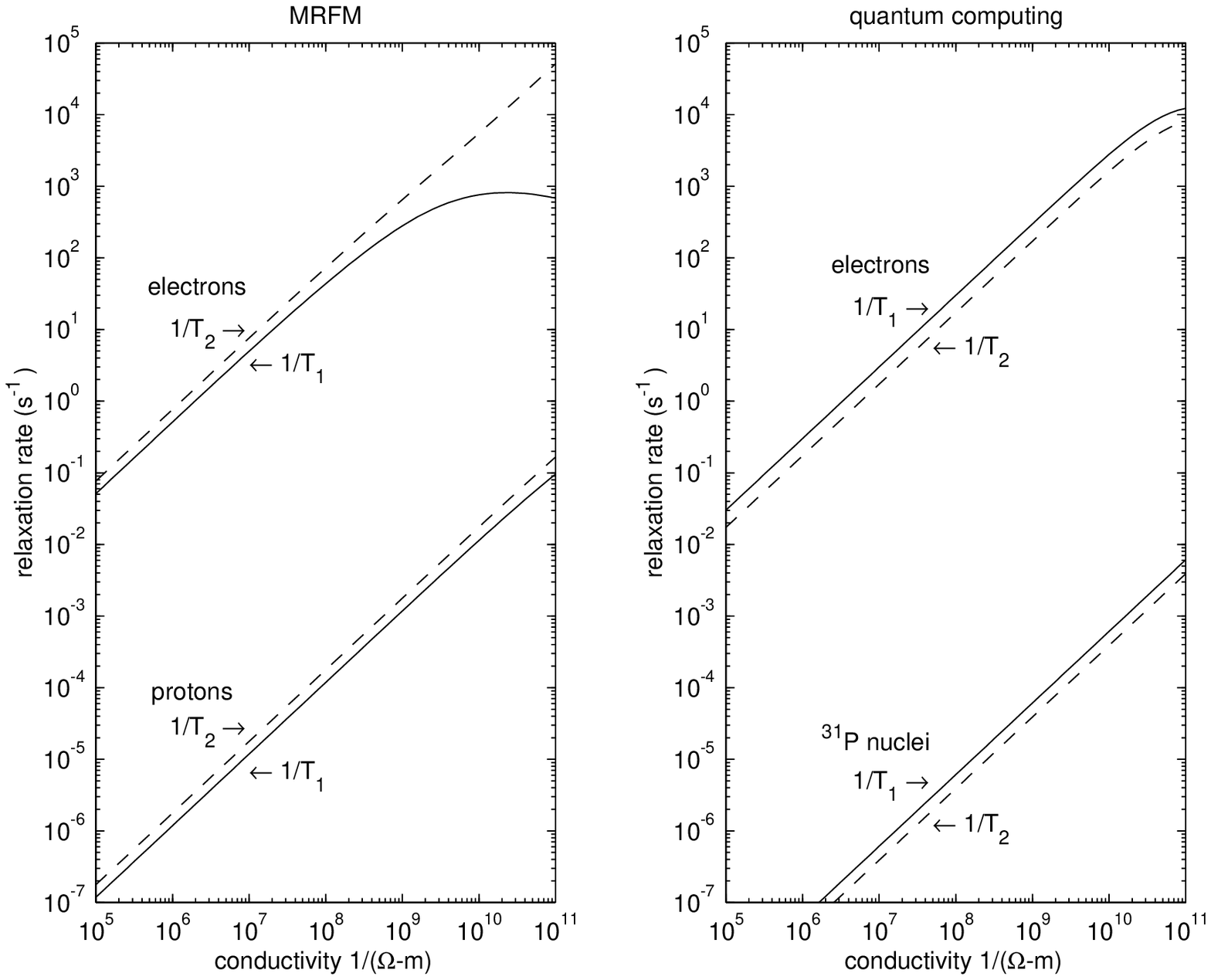} 
	\caption{Predicted spin 
	relaxation rates in magnetic resonance force microscopy and in 
	quantum computing, as a function of conductivity.}
 	\label{fig:Relaxation}
\end{center}
\end{figure}

\end{document}

%% file: MRFM_Magnetic_Part_01.tex
\section{Introduction}

\noindent The engineering control of quantum decoherence is central to
several emerging scientific goals.  One such goal is the direct
observation of molecular structure---in three dimensions and with
single-atom resolution---by magnetic resonance force microscopy
\cite{Sidles:95}.  Another is the solution of otherwise intractable
mathematical problems---like factoring large numbers---by quantum
computing \cite{Schor:97}.

At present, the main practical obstacle to achieving these goals is
that the various mechanisms that cause quantum decoherence are not yet
fully cataloged and understood.  In consequence, quantum-coherent
experiments often reveal unexpectedly fast decoherence rates, due to
unanticipated---or even previously unknown---relaxation mechanisms. 
Similar struggles with decoherence have occupied previous scientific
generations; Orbach \cite{Orbach:72} for example reviews the
forty-year struggle to achieve a reasonably comprehensive
understanding of relaxation in electron spin resonance.

This article is concerned with a decoherence mechanism which is
ubiquitous in quantum-coherent technologies: \emph{thermal magnetic
noise}.  Physically speaking, thermal magnetic noise is created by the
same thermal current fluctuations that create Johnson noise.  Thus,
thermal magnetic noise is present in any device that contains electric
conductors.

We will be mainly concerned with practical engineering aspects of 
thermal magnetic noise, but we will also give an explicit 
example---the first in the literature to our knowledge---of a 
\emph{fluctuation-dissipation-entanglement} theorem.  Such theorems 
may be broadly defined as invertible functional relations between the 
dissipative kernel of a system and the system's quantum entanglement 
with a thermal reservoir, such that the entanglement determines the 
dissipation and \emph{vice versa}.  

This article is organized as follows.  Prior work relating to thermal
magnetic noise is reviewed in Section~\ref{Section: Background}.  The
main new results are summarized in Section~\ref{Section: Physical
motivation and main results}, these results consist of closed-form
expressions for the magnetic noise spectral density
(Section~\ref{sec:ThermalMagneticSpectralDensities}), quantum
decoherence (Section~\ref{Section: Quantum decoherence}), and a
fluctuation-dissipation-entanglement theorem (Section~\ref{Section: A
fluctuation-dissipation-entanglement theorem}).  The practical
implications of these results for quantum-coherent engineering are
discussed in Section~\ref{Section: Quantum engineering implications}
via worked examples for trapped atom experiments
(Section~\ref{Section: Atom and ion traps}), magnetic resonance force
microscopy (Section~\ref{Section: Magnetic resonance force
microscopy}), and quantum computing (Section~\ref{Section: Quantum
computing}).  Some topics for further research are suggested in
Section~\ref{Section: Conclusions}.  

Algebraically tedious derivations are relegated to three appendices. 
Appendix~\ref{Appendix: Fluctuation and dissipation} derives a
closed-form expression for the magnetic noise spectral density,
Appendix~\ref{Appendix: Quantum decoherence} derives quantum Langevin
and Bloch equations that describe decoherence in the presence of
thermal magnetic noise, and Appendix~\ref{Appendix: A
fluctuation-dissipation-entanglement theorem} derives
fluctuation-dissipation-entanglement theorems for both
spin-$\tfrac{1}{2}$ particles and harmonic oscillators, as well as
relations between dissipation and renormalization.
%

\subsection{Prior work relating to thermal magnetic noise}
\label{Section: Background}

The quantum field theory of electromagnetic fluctuations in lossy
linear media is presented in textbooks by Landau, Lifshitz, and
Piaevskii \cite{Landau:80Thermal} and by Rytov \emph{et al.}
\cite{Rytov:89}.  Far from a warm body, this theory describes
fluctuations which are simply the familiar phenomenon of black-body
radiation.  Near to a warm body, the situation is considerably more
complex.  Non-radiating terms in the field theory become important,
and create phenomena such as the Casimir force (identical to the Van
der Waals force), which is familiar from chemistry and force
microscopy.  Recently, Dorofeyev \emph{et al.} \cite{Dorofeyev:99}
have observed the attractive, fluctuating and dissipative components
of the Casimir force, which were in good agreement with
field-theoretic predictions.

The Casimir force arises mainly from fluctuating electric fields,
which do not directly couple to the spin magnetic moments.  Our
investigation instead focuses on thermal fluctuations in magnetic
fields.

To the best of our knowledge, the theory and experiments of Varpula
and Poutanen \cite{Varpula:84}, as reviewed and extended by Nenonen,
Montonen, and Katila \cite{Nenonen:96}, were the first experimental
observation and theoretical analysis of thermal magnetic fluctuations. 
Their work focussed on biomagnetism experiments, in which magnetic
fluctuations originate in the copper walls of shielded rooms; such
noise can be easily large enough to obscure biomagnetic signals.  They
presented a phenomenological model of thermal magnetic noise in which
metallic conductors were modeled as collections of independent
thermally excited resistive elements.  The resulting predictions were
in excellent accord with experiment.

Quantum-coherent technologies operate in a vastly different regime
from biomagnetism experiments.  Ambient temperatures $T$ are as low as
is experimentally feasible---a few Kelvins down to millikelvins---and
observation frequencies $\omega$ are megahertz to gigahertz, such that
$\hbar\omega\gg k_{\mathrm B}T$.

Quantum-coherent phenomena dominate this physical regime.  The
applicability of the Varpula-Poutanen noise model then becomes
uncertain, because the model is derived from a purely phenomenological
description of thermal noise in room-temperature conductors.
We have therefore sought to improve and extend the Varpula-Poutanen
model in five respects:
\begin{enumerate}
    
	\item Rigorous quantum mechanical and thermodynamic foundations
	have been provided via the fluctuation-dissipation theorem.

	\item The results now encompass superconducting materials and/or
	materials with non-vanishing magnetic susceptibility.
	
	\item A closed-form expression for the thermal magnetic noise
	spectral density has been obtained.
	
	\item Bloch equations describing quantum decoherence have been derived 
	via a quantum Langevin formalism and the independent oscillator 
	heat bath model of Ford, Lewis, and O'Connell\cite{Ford:88}.

	\item Fluctuation-dissipation-entanglement theorems have been
	constructed, and the link between dissipation and renormalization
	has been clarified.

\end{enumerate}
In carrying through this program, we have been able to prove that the
Varpula-Poutanen model is \emph{exact} in the physical regime relevant
to biomagnetism: real conductivity, macroscopic length scales, and
large temperature.  This explains why the Varpula-Poutanen model
yields predictions which are in excellent accord with experimental
measurements of biomagnetism~\cite{Varpula:84}, and is a tribute to
their pioneering physical insight.

\section{Physical motivation and main results}
\label{Section: Physical motivation and main results}

The necessary existence of thermal magnetic noise can be deduced by
considering a loop antenna held near a conducting slab, as shown in
Fig.~\ref{fig:SkinCoil}.  If a voltmeter is placed across the
terminals of such an antenna, a fluctuating voltage $V(t)$ will be
observed.  The spectral density\footnote{Our normalization for
spectral densities is $S_{V}(\omega) = \int_{-\infty}^{\infty}d\tau\,
e^{-i\omega\tau}\expect{V(0)V(\tau)}$, where $\expect{V(0)V(\tau)}$ is
the voltage autocorrelation.  Thus
$\expect{V^{2}}=1/(2\pi)\,\int_{-\infty}^{\infty}d\omega\,
S_{V}(\omega)$, so that spectral densities are two-sided (positive and
negative frequencies), with bandwidths given in Hertz.}
$S_{V}(\omega)$ of this \emph{Johnson noise} is related to the complex
antenna impedance $Z(\omega)$ by the well-known equation
\cite{Landau:80Thermal}
\begin{align}
\label{eq:RNoise}
S_{V}(\omega) &=
	 \text{Re}\big(Z(\omega)\big)\ \hbar\omega \coth\left(\dfrac{\hbar\omega}
	 {2k_{\text{B}}T}\right), &
	 	& \text{in general;}\\ \notag
	 &\simeq \text{Re}\big(Z(\omega)\big)\ 2 k_{\text{B}} T,& &
	 	\text{for $k_{\text{B}}T\gg\hbar\omega$.}
\end{align}
Here $\text{Re}\big(Z(\omega)\big)$ is the resistive 
impedance.  This result is just the fluctuation-dissipation 
theorem as it applies to fluctuating 
voltages in dissipative impedances.

Now we ask, what is the physical origin of these voltage fluctuations?  
We suppose that the coil has negligible intrinsic resistance, in which 
case the voltage can only have been induced by a magnetic flux 
threading the antenna loop.  In turn, this magnetic flux can only have 
been generated by thermally excited currents in the nearby conducting 
slab.  

So wherever Johnson noise is present, thermal magnetic noise will be
present also, because the same underlying current fluctuations
generate both phenomena.
%

\subsection{Thermal magnetic spectral densities}
\label{sec:ThermalMagneticSpectralDensities}
We will begin by presenting closed-form expressions for the spectral
density of thermal magnetic noise, and discussing their physical
significance.  Deriving these expressions is tedious, and is
deferred until Appendix~\ref{Appendix: Fluctuation and dissipation}.

Since the thermal magnetic field $\lb{B}(t)$ is a vector, its 
autocorrelation $\lb{C_{B}}(\tau)$ and spectral density 
$\lb{S_{B}}(\omega)$ are matrix-valued functions defined by
\begin{align}  
	\lb{C_{B}}(\tau) &= 
	\expect{\lb{B}(0)\gsb{\otimes}\lb{B}(\tau)},\\ 
	\lb{S_{B}}(\omega) & =  \int_{-\infty}^{\infty}
	\!e^{-i\omega\tau}\,\lb{C_{B}}(\tau)\,d\tau
\intertext{where $\gsb{\otimes}$ denotes the outer product of two vectors:
$(\lb{a}\gsb{\otimes}\lb{b})_{ij} \equiv a_{i}b_{j}$.  
As shown in Section~\ref{sec:SolvingTheMaxwellEquations}, 
$\lb{S_{B}}(\omega)$ is of the general form:}
\label{eq:GammaNoise}
\lb{S_{B}}(\omega) & = \gsb\Gamma(\omega)
	 \hbar\omega \coth\left(\dfrac{\hbar\omega}{2k_{\text{B}}T}\right)\\
	 \notag
& = \big[\lb{I}+\lbhat{n}\gsb\otimes\lbhat{n}\big] \,\Gamma(\omega)
	 \hbar\omega \coth\left(\dfrac{\hbar\omega}{2k_{\text{B}}T}\right)
\end{align}
where $\lbhat{n}$ is the unit vector normal to the slab surface and 
\lb{I} is the 3$\times$3 identity matrix.  

The \emph{magnetic dissipation tensor} $\gsb\Gamma(\omega)$ in 
\eqref{eq:GammaNoise} plays the same fundamental role that resistance 
plays in Johnson noise, in the sense that---as we will see---knowledge 
of $\gsb\Gamma(\omega)$ quantitatively determines a broad spectrum of 
fluctuation, dissipation, and quantum entanglement phenomena.

\subsubsection{Device parameters}
\label{Section: Device parameters}

In general, we will concerned with thermal magnetic noise induced at a 
distance $d$ from a conductive slab of thickness $t$, as illustrated 
in Fig.~\ref{fig:Regions}.  The slab is assumed to have a 
frequency-dependent complex conductivity $\sigma(\omega)$ and 
permeability $\mu(\omega)$; for compactness we suppress the 
frequency-dependence of $\sigma$ and $\mu$.  All results are presented 
in S.I.\ units, and all time-dependence is assumed to be $e^{i\omega 
t}$.  See Section~\ref{sec:SolvingTheMaxwellEquations} for a statement 
of Maxwell's equations in this convention.

The four device parameters $\{d,t,\sigma,\mu\}$ may be regarded as 
fundamental; all other parameters are derived from them.  The single 
most important derived parameter is the \emph{skin depth} $\lambda 
\equiv |\omega\mu\sigma|^{-1/2}$.  Another important derived parameter 
is the phase $\phi$ of the complex conductivity, as defined by 
$\lb{j}=\sigma \lb{E} = |\sigma| e^{-i\phi}$, where $\lb{E}$ is the 
E-field and $\lb{j}$ is the current. 

\subsubsection{Superconductors}
For the special case of an ideal superconductor, the London equation 
can be stated in the form $i \mu_{0} \omega \lb{j} = 
\lambda_{\text{\tiny L}}^{2}\lb{E}$, where $\lambda_{\text{\tiny L}}$ is the 
\emph{London penetration depth}.  Comparing this with the definition 
of the conductivity $\lb{j}=\sigma \lb{E}$, we see that for an ideal 
superconductor $\sigma = -i \lambda_{\text{\tiny L}}^{2}/(\mu_{0}\omega)$, 
$\lambda = \lambda_{\text{\tiny L}}$, and $\phi=\pi/2$.  Thus ideal 
superconductors do not have infinite conductivity, but rather have 
a wholly imaginary conductivity.

Real superconductors approximate ideal superconductors only at zero 
temperature.  At finite temperatures a mixture of superconducting and 
normal phase charge carriers is present, such that the conductivity 
has an intermediate phase $\phi\in(0,\pi/2)$.  See \cite{Bonn:96} for 
a review of the literature, including measurements of $\sigma$ at 
finite temperature.

%

\subsubsection{Single paramagnetic or diamagnetic conducting slabs}
Most nonferrous metals and insulators have relative magnetic 
permeability $K\equiv\mu/\mu_{0}\simeq1$, where $\mu$ is the 
permeability of the material (the slab in our case), and $\mu_{0}$ is 
the permeability of the vacuum.  In Appendix~\ref{Appendix: 
Fluctuation and dissipation} we derive the following asymptotic 
dependance of the magnetic dissipation for $K=1$
\begin{equation}
\label{eq:GammaResults}
\Gamma(\omega) = \dfrac{\mu_{0}^{2}\,\text{Re}(\sigma)}{64\pi}\times
	\begin{cases}
	\dfrac{t}{d(t+d)},
		&\lambda\gg\text{min}\{d,\sqrt{dt}\};\\[2.5ex]
	\dfrac{3\lambda^{3}}{d^{4}\,\cos(\pi/4-\phi/2)},	
		&\lambda\ll\text{min}\{d,t\};\\[2.5ex]
	\dfrac{2\lambda^{4}(dt-8\lambda^{2}\sin\phi)}{d^{5}t^{2}}.&
	t\ll\lambda\ll\sqrt{dt}.
	\end{cases}
\end{equation}
Collectively, these three limits span the entire 
$\{d,t,\sigma,\omega\}$ design space.  The following expression 
smoothly interpolates all three limits:
\begin{equation}
\label{eq:GammaApproximation}
\Gamma(\omega) \simeq\  
	\frac{3 \mu_{0}^{2}\,\text{Re}(\sigma)\lambda^{3} t \cos\phi}
	{64\pi\big(3\lambda^{3}d(d+t)+ (1-e^{-\alpha_{\text{c}}}) t d^{2}(d+2\lambda)^{2}
	\cos(\pi/4-\phi/2)\big)}.
\end{equation}
It is constructed by adding the first and second asymptotic limits of 
\eqref{eq:GammaResults} inversely, then introducing a transition 
parameter 
\begin{equation}
\alpha_{\text{c}}=(d t+ 8 \lambda^{2} \sin\phi)/(2 d 
\lambda \cos(\pi/4-\phi/2)) 
\end{equation}
such that the third asymptotic limit also is smoothly interpolated.

A~numerical survey shows that the resulting closed-form expression 
\eqref{eq:GammaApproximation} predicts noise levels that are no more 
than $1.75\text{ dB}$ too large and $0.5\text{ dB}$ too small 
everywhere in $\{d,t,\sigma,\omega\}$ design space---an accuracy which 
suffices for almost all practical design work.

\subsubsection{The case of two conducting slabs} 
The case of magnetic fluctuations measured at a point midway between 
two identical conducting slabs is also solved in 
Appendix~\ref{Appendix: Fluctuation and dissipation}, and the resulting 
asymptotic expressions are precisely double those given in 
\eqref{eq:GammaResults}.  A numerical survey shows that doubling the 
one-slab noise yields a slight overestimate of the two-slab midpoint 
noise, but the error is no more than $0.6\text{~dB}$ for all 
$\{d,t,\sigma,\omega\}$.  Thus, to a very good approximation two 
adjacent slabs each generate independent noise at the midpoint.


\subsubsection{The general case} 
In the general case in which both lossy magnetic permeability and 
electric conductivity are combined with a skin depth $\lambda$ 
comparable to $d$, no analytic results are presently available.  
Appendix~\ref{Appendix: Fluctuation and dissipation} presents a 
one-dimensional integral \eqref{eq:integrand} which must be 
numerically evaluated to determine the magnetic spectral densities.  

\subsubsection{Comparison with experiment} 

Although our general integral expression \eqref{eq:integrand} is 
superficially of a different form than the Varpula-Poutanen model 
\cite{Varpula:84}, and is derived from a very different underlying 
model, a numerical comparison shows that these two models yield 
identical predictions in the physical regime relevant to biomagnetic 
experiments: real conductivity, audio frequencies, vanishing magnetic 
susceptibility, and large temperature.  Since in this regime the 
Varpula-Poutanen model has been shown to agree very well with 
experiment, we may regard the model presented in this article as 
having been at least partially validated.  

More extensive comparisons of our model with the Varpula-Poutanen 
model are not feasible, because the V-P model does not contain 
parameters corresponding to the magnetic susceptibility $\mu$, 
conductivity phase $\phi$, or the quantum of action $\hbar$.

We note also that \emph{no} model of thermal magnetic noise has yet 
been experimentally tested at microscopic scales, low temperatures, and high 
frequencies, in materials that are wholly or partially 
superconducting, or are magnetic---which is precisely the regime of 
greatest importance to quantum-coherent engineering.

\subsection{Quantum decoherence}
\label{Section: Quantum decoherence}
Now we turn our attention to the quantum decoherence induced by 
thermal magnetic fields.  We assume a two-state system coupled to 
the magnetic thermal field $\lb{B}(t)$ via the Hamiltonian
\[
	H = -\gamma \big(\lb{B}_{0}+\lb{B}(t)\big)\gsb\cdot\lb{s}
\]
where $\gamma$ is the gyromagnetic ratio of the spin, 
$\lb{B}_{0}$ is a constant polarizing field, and the spin angular 
momentum $\lb{s}$ satisfies the usual commutation relations
$[s_{i},s_{j}]=i\hbar\epsilon_{ijk}\,s_{k}$.

Our results will be expressed in terms of the rotating-frame 
magnetic spectral density $\lb{S}_{\lb{B}_{\text{rot}}}(\omega)$, 
which is given in terms of the laboratory-frame density 
$\lb{S}_{\lb{B}}(\omega)$ \eqref{eq:GammaNoise} via
\begin{multline}
\label{eq:RotatingSpectralDensity}
\lb{S}_{\lb{B}_{\text{rot}}}(\omega) =
	\big(\lbhat{b}\gsb\otimes\lbhat{b}\big)\ 
	\text{tr}\big[\lbhat{b}\gsb\otimes\lbhat{b}
	\gsb\cdot\lb{S}_{\lb{B}}(\omega)\big]+\\ 
	 \qquad\big(\lb{I}- \lbhat{b}\gsb\otimes\lbhat{b}\big)\ \frac{1}{2}	
	\,\text{tr}\big[(\lb{I}- 
\lbhat{b}\gsb\otimes\lbhat{b})
\gsb\cdot\lb{S}_{\lb{B}}(\omega_{0})\big],
\end{multline}
where the precession frequency $\omega_{0}$ about the unit axis 
of spin precession $\lbhat{b}$ is given by 
$\gamma\lb{B}_{0}\equiv\omega_{0}\lbhat{b}$, and 
$\text{tr}[\ldots]$ is the matrix trace (the sum of diagonal 
elements).

In Appendix~\ref{Appendix: Quantum decoherence} we derive the quantum 
Langevin equations for a spin-$\tfrac{1}{2}$ particle coupled to a 
thermal magnetic field, and from them we show that the time evolution 
of the spin density matrix is described by Bloch-type equations with 
relaxation times $T_{1}$, $T_{2}$, and $T_{1\rho}$ given by
\begin{subequations}
\label{eq:relaxations}
\begin{align}
\label{eq:G1covariant}
\frac{1}{T_{1}} &= \frac{1}{2}\,\gamma^{2}\,\text{tr}\big[(\lb{I}- 
\lbhat{b}\gsb\otimes\lbhat{b})
\gsb\cdot\lb{S}_{\lb{B}_{\text{rot}}}(0)\big]\\
\label{eq:G2covariant}
\frac{1}{T_{2}} &= \frac{1}{2T_{1}}+ \frac{1}{2}\gamma^{2}\,
\text{tr}\big[\lbhat{b}\gsb\otimes\lbhat{b}\gsb\cdot
\lb{S}_{\lb{B}_{\text{rot}}}(0)\big]\\
\frac{1}{T_{1\rho}} &= \frac{1}{2}\,\gamma^{2}\,\text{tr}\big[(\lb{I}- 
\label{eq:G1rhocovariant}
\lbhat{b}_{1}\gsb\otimes\lbhat{b}_{1})
\gsb\cdot\lb{S}_{\lb{B}_{\text{rot}}}(\omega_{1})\big]
\end{align}
\end{subequations}
Here $T_{1\rho}$ is the relaxation time in the presence of a 
radio-frequency (RF) applied field---commonly called a ``spin-locking'' 
field.  In the frame co-rotating with the RF field, it appears as a 
constant B-field $\lb{B}_{1}=B_{1}\lbhat{b}_{1}$ whose characteristic
precession frequency is $\omega_{1}=\gamma B_{1}$.  Substituting 
(\ref{eq:GammaNoise}--\ref{eq:RotatingSpectralDensity}) in 
(\ref{eq:G1covariant}--\ref{eq:G2covariant}), we obtain 
$\{T_{1},T_{2},T_{1\rho}\}$ explicitly in terms of the magnetic 
damping coefficient $\Gamma(\omega)$ given in 
\eqref{eq:GammaApproximation}
\begin{subequations}
\label{eq:G1andG2expanded}
\begin{align}
\label{eq:G1expanded}
\frac{1}{T_{1}} &= 
\frac{1}{2}\,\gamma^{2}\,(3-\cos^{2}\theta)\,\Gamma(\omega_{0})\,
\hbar\omega_{0}\,\coth\left(\frac{\hbar\omega_{0}}{2 k_{\text{B}}T}\right)\\
\label{eq:G2expanded}
\frac{1}{T_{2}} &=  
\frac{1}{2T_{1}}+
\frac{1}{2}\gamma^{2} (1+\cos^{2}\theta)\,\Gamma(0)\,2k_{\text{B}}T\\
\label{eq:G3expanded}
\frac{1}{T_{1\rho}} &=  
\frac{1+\cos^{2}\beta}{2T_{1}}+\\
\nonumber
&\quad\quad\quad \frac{1}{2}\gamma^{2} \sin^{2}\!\beta\ (1+\cos^{2}\theta)\,\Gamma(\omega_{1})\,
\hbar\omega_{1}\,\coth\left(\frac{\hbar\omega_{1}}{2 k_{\text{B}}T}\right)
\end{align}
\end{subequations}
Here $\cos\theta=\lbhat{b}\gsb\cdot\lbhat{n}$ and 
$\cos\beta=\lbhat{b}\gsb\cdot\lbhat{b}_{1}$, where 
$\lbhat{b}_{1}$ is the RF field axis in the rotating frame, 
\lbhat{b} is the polarization axis, and \lbhat{n} is the unit 
vector normal to the slab surface.

Note that $\Gamma(\omega)$ appears with varying frequency 
arguments $\omega\in\{0,\omega_{1},\omega_{0}\}$; this has 
important design consequences because in most cases 
$\Gamma(\omega)$ exhibits a strong frequency dependence.

As far as practical quantum engineering is concerned,  
expressions (\ref{eq:G1expanded}--\ref{eq:G3expanded})
are the main ``deliverable'' of this article. 

\subsection{A fluctuation-dissipation-entanglement theorem}
\label{Section: A fluctuation-dissipation-entanglement theorem}
From a purely formal point of view, 
fluctuation-dissipation-entanglement theorems exist for a simple 
reason: the same Hamiltonian matrix elements that control fluctuation 
and dissipation also control certain measures of quantum entanglement; 
Appendix~\ref{Appendix: A fluctuation-dissipation-entanglement 
theorem} discusses this point of view.

Physically speaking, we reason as follows.  We consider a two-state
quantum system, which---as in the real world---interacts weakly with a
thermal reservoir.  We adjust the temperature of the reservoir to
zero, and ask: what is the probability that the two-state system is
\emph{not} in its ground state?  To the extent that this probability
is non-zero, it describes an irreducible quantum entanglement with the
thermal reservoir.

Now, at zero temperature the classical entanglement probability is 
zero, and even the Bloch equations \eqref{eq:FullBlochEquations}, 
which have a firmer quantum justification, relax spin systems to their 
ground state at zero temperature, and thus predict zero 
entanglement.

However, a higher-order calculation reveals that the entanglement 
probability is finite even at zero temperature.  We calculate this 
probability as follows.  As before, we consider a spin-$\tfrac{1}{2}$ 
particle magnetically coupled to a thermal reservoir.  We polarize the 
spin via an external field $\lb{B}_{0}$, which lifts the degeneracy of 
the spin system along a polarization axis $\omega_{0}\lbhat{p} = 
\gamma \lb{B}_{0}$, with $\gamma$ the gyromagnetic ratio of the spin 
and $\omega_{0}$ the (by convention positive) precession frequency.  
We let $\ket{\psi_{0}}$ be the ground state of the isolated spin, and 
we define a projection operator $P_{\lcal{E}} = \lbb{I} - 
\ket{\psi_{0}}\gsb\otimes \bra{\psi_{0}}$, with $\lbb{I}$ the identity 
operator.  By construction, the expectation value of $P_{\lcal{E}}$ 
vanishes for an isolated ground state: 
$\braket{\psi_{0}}{P_{\lcal{E}}}{\psi_{0}} = 0$.

Of course, no real-world spin system is perfectly isolated.\footnote{ 
If nothing else, the spin can exchange quanta with the vacuum, which 
may be regarded as a zero-temperature thermal reservoir.  It follows 
that fluctuation-dissipation-entanglement theorems can be constructed 
in quantum field theories, where their significance remains to be 
elucidated.} We therefore enlarge our Hilbert space to encompass a 
thermal reservoir with a potentially infinite set of basis states $\{ 
\ket{\phi_{i}}; i\in 0,1,\ldots,\infty\}$.  We generalize 
$P_{\lcal{E}}$ to include the thermal reservoir by defining a new 
projection operator $\lcal{E}$
\begin{equation}
	\label{eq:defOfE}
	\lcal{E} \equiv \sum_{i} \ket{\phi_{i}}\otimes P_{\lcal{E}}\otimes 
	\bra{\phi_{i}}
\end{equation}
which projects onto the subspace in which the spin is 
\emph{not} in its ground state.

It is a well-defined problem to calculate, perturbatively, the 
expectation $\braket{\Psi_{0}}{\lcal{E}}{\Psi_{0}}$, where 
$\ket{\Psi_{0}}$ the ground state of the combined 
spin-plus-thermal-reservoir system.  Keeping in mind that $\lcal{E}$ 
and $\ket{\Psi_{0}}$ depend on $\lbhat{p}$ and $\omega_{0}$, we can define 
a scalar entanglement function $E(\lbhat{p},\omega_{0})\equiv 
\braket{\Psi_{0}}{\lcal{E}}{\Psi_{0}}$.  In Appendix~\ref{Appendix: A 
fluctuation-dissipation-entanglement theorem} we show that 
$E(\lbhat{p},\omega_{0})$ is related to the magnetic dissipation 
tensor $\gsb\Gamma(\omega)$ by the following 
\emph{fluctuation-dissipation-entanglement theorem:}
\begin{align}
	\label{eq:fde theorem}
	E(\lbhat{p},\omega_{0}) 
& = \frac{\gamma^{2}\hbar}{4 \pi}\ 
\text{\large $\lfrak{G}$}_{2}\big\{\omega\,\text{tr}\big[
(\lb{I}-\lbhat{p}\gsb\otimes\lbhat{p})
\gsb\cdot\gsb\Gamma(\omega)\big];\omega_{0}\big\}
\end{align}
Here $\lfrak{G}_{2}$ is a \emph{\Stieltjies transform}, in Bateman's 
notation \cite{Bateman:54}, as defined by
\begin{equation}
   \text{\large $\lfrak{G}$}_{\!\rho}\{f(x);y\} = \int_0^{\infty}\!\!dx\ 
                f(x) (x+y)^{-\rho} 
\end{equation}
We note that \Stieltjies transforms are invertible, such that a 
measurement of the entanglement function $E(\lbhat{p},\omega_{0})$ 
suffices in principle to determine the magnetic dissipation tensor 
$\gsb{\Gamma}(\omega)$ and \emph{vice versa}.

For practical quantum engineering purposes, the main utility of this 
theorem is that it allows us to determine the hard-to-measure 
entanglement from the easier-to-measure (or predict) dissipation.  For 
the particular case of thermal magnetic noise, the frequency 
dependence (\ref{eq:GammaApproximation}) of $\gsb{\Gamma}(\omega)$ is 
such that the \Stieltjies integral is absolutely convergent, so that 
the entanglement of real-world devices can be readily be calculated.

\subsubsection{Approximate expression for the entanglement}
\label{Section: Approximate expression for the entanglement}

In general, the \Stieltjies integral \eqref{eq:fde theorem} cannot be
evaluated in closed form.  However, a simple and physically meaningful
approximate evaluation is possible.  We begin by remarking that
dissipative kernels always have a cut-off frequency
$\omega_{\text{c}}$, because otherwise their associated noise spectral
density would carry infinite power.  For the particular case of
thermal magnetic noise originating in a thick plate, the cut-off
frequency is such that $\lambda(\omega_{\text{c}}) \sim d$.  We assume
the frequency of interest $\omega_{0}$ is small compared to the
cut-off frequency, so that $\omega_{\text{c}}\gg\omega_{0}$, and we
brutally approximate $\gsb{\Gamma}(\omega)$ as constant over the range
$\omega\in(\omega_{0},\omega_{\text{c}})$, and zero elsewhere.  Then
the \Stieltjies integral \eqref{eq:fde theorem} can be evaluated in
closed form as
\begin{align}
	\label{eq:approx entanglement}
	E(\lbhat{p},\omega_{0}) & \simeq 
	\frac{\gamma^{2}\hbar}{4 \pi}\,
	\text{tr}
	\big[ (\lb{I}-\lbhat{p}\gsb\otimes\lbhat{p})
	\gsb\cdot\gsb\Gamma(\omega_{0})\big]\,
	\ln(\omega_{\text{c}}/\omega_{0})
	& & \text{from \eqref{eq:fde theorem}}\\\notag
	& = \frac{\gamma^{2}\hbar}{4 \pi}\,
	\text{tr}
	\big[ (\lb{I}-\lbhat{p}\gsb\otimes\lbhat{p})
	\gsb\cdot\frac{\lb{S}_{\lb{B}}(\omega_{0})}{\hbar\omega_{0}}\big]\,
	\ln(\omega_{\text{c}}/\omega_{0})
	& & \text{by \eqref{eq:GammaNoise} for $T\to 0$}\\ \notag
	&=\frac{\ln(\omega_{\text{c}}/\omega_{0})}
	{2 \pi \omega_{0}T_{1}}
	& & \text{by \eqref{eq:RotatingSpectralDensity} and \eqref{eq:G1covariant}}
\end{align}	
where $T_{1}$ is evaluated at zero temperature.

\section{Quantum engineering implications}
\label{Section: Quantum engineering implications}
We will now apply (\ref{eq:GammaNoise}--\ref{eq:G1andG2expanded}) in 
the design analysis of representative quantum technologies.  Our goal 
is partly to illustrate practical applications of our results, and 
partly to identify areas where further research is needed.

A small but important point: when quoting numerical values $\propto 1/ 
\sqrt{\text{\small Hz}}$ we shall embrace the usual engineering 
convention that bandwidths encompass only positive frequencies; this 
requires the insertion of an additional factor of two when evaluating 
two-sided spectral densities, \emph{e.g.}~\eqref{eq:GammaNoise}.

\subsection{Atom and ion traps} 
\label{Section: Atom and ion traps} 

We begin by considering thermal magnetic noise at centimeter length 
scales and audio frequencies, at room temperature.  This is the same 
regime considered by Varpula and Poutanen 
\cite{Varpula:84,Nenonen:96}, and it is also a regime typical of at 
least some atomic physics experiments \cite{Jacobs:95}.  

Specifically, we will calculate the spectral density of the thermal 
magnetic fields between room-temperature two copper slabs, each $t=1\ 
\text{cm}$ thick, spaced $2d=2\ \text{cm}$ apart.  We are particularly 
interested in the zero-frequency spectral density in the \lbhat{n} 
direction normal to the slab surface, because this describes thermally 
induced fluctuations in the background polarizing field.  At room 
temperature $T=300\ \text{K}$ the conductivity of high-purity copper 
is of order $5.9\times10^{7}\ (\Omega\text{\ m})^{-1}$.  From 
\eqref{eq:GammaNoise} and \eqref{eq:GammaApproximation}, with $d=t=1\ 
\text{cm}$ and $\phi=0$, and taking into account that we have two 
plates whose noise is additive, we find at zero frequency 
$(\lbhat{n}\gsb\cdot\lb{S}_{\lb{B}}(0\,\text{Hz})\gsb\cdot\lbhat{n})^{1/2} 
\sim 1.2\ \text{pT}/\sqrt{\text{\small Hz}}$, rolling off to $ 
(\lbhat{n}\gsb\cdot\lb{S}_{\lb{B}}(100\,\text{Hz})\gsb\cdot\lbhat{n})^{1/2}\sim 
0.6\ \text{pT}/\sqrt{\text{\small Hz}}$ at 100 Hertz.

Until recently, such picoTesla fluctuations would have been viewed as 
being of no practical consequence.  However, magnetic fields change 
sign under time reversal, and hence magnetic fluctuations locally 
violate time-reversal invariance, and so these fluctuations must be 
understood and controlled in high-precision tests of fundamental 
physics.

For example, the most stringent experimental limit on the electric 
dipole moment $d_{\text{\tiny E}}$ of a fundamental particle is 
$d_{\text{\tiny E}} \le 8.7 \times 10^{-28}\ \text{$e$ cm}$ for the 
mercury isotope $\oneninenineHg$ \cite{Jacobs:95} (here $e$ is the 
electron charge).  Typically, dipole moments are measured by observing 
the change in precession frequency induced by an applied electric 
field of magnitude $E$.  It follows that $\lb{S}_{\lb{B}}$ can be 
expressed as an equivalent dipole noise spectral density $S_{d} = 
(\hbar\gamma/(2 
E))^{2}\,\lbhat{n}\gsb\cdot\lb{S}_{\lb{B}}(0)\gsb\cdot\lbhat{n}$.  For 
$\oneninenineHg$, the gyromagnetic ratio $\gamma/(2\pi) = 7.59\ 
\text{MHz/Tesla}$.  This yields a zero-frequency equivalent dipole 
noise for our example of room-temperature copper plates of
\[ 
ES_{d}^{1/2} \sim 1.94 \times 10^{-20}\ \text{eV}/
\sqrt{\text{\small Hz}}.
\]
For a typical electric field $E\sim10^{4}\ \text{V/cm}$, the 
equivalent dipole noise would therefore be $S_{d}^{1/2}\sim1.94 
\times 10^{-24}\ \text{$e$ cm}/\sqrt{\text{\small Hz}}$.

This is a substantial noise level: it would naively require an 
averaging time on the order of $10^{7}$ seconds to better the 
published electric dipole moment limit \cite{Jacobs:95}.  And it would 
not be entirely straightforward to reduce the thermal magnetic noise 
by making the copper shields thinner or less conductive, because this 
would defeat their purpose of shielding the experiment from external 
fields.

Fortunately, another effect provides mitigation: trapped atom
experiments typically measure the net signal from atoms in many
different regions of a cell.  To the extent that $n$ independent
regions are averaged, the equivalent dipole noise power will be
reduced by a factor of $1/n$.  Some of the formalism necessary for
calculating $n$ is set forth in Appendix~\ref{Section: Spatial
correlations and gradients}.  Detailed calculations would be
strongly dependent on the particular design chosen.

\subsection{Magnetic resonance force microscopy}
\label{Section: Magnetic resonance force microscopy}

Magnetic resonance force microscopy (MRFM) is a quantum-coherent
technology whose ultimate objective is to produce magnetic resonance
images of individual molecules \emph{in situ}, nondestructively, in
three dimensions, with Angstrom resolution.  Such a technology would
allow much of molecular biology to be conducted as an observational
science---along the lines of, \emph{e.g.}, astronomy---rather than an
experimental science.

We remark that a single-spin MRFM imaging device can be alternatively
regarded as a first-generation solid-state quantum computer, in which
the individual qubits carry binary information about the presence or
absence of a spin spin at a specified atomic coordinate.

There are two main experimental challenges in MRFM. The first is that
single-spin signal forces are exceedingly small, of order
$10^{-18}\,\text{N}$ for electron magnetic moments.  The MRFM
community's design options for achieving the required sensitivity are
reasonably well understood (but challenging to implement in practice):
reduce the mass of the cantilever, increase its damping time, reduce
the ambient temperature, and employ a sharper magnetic tip with a
higher field gradient.

The second---emerging---challenge in MRFM is that the spin state must
maintain its quantum coherence for a time long enough to detect the
signal force.  Here the design issues are \emph{not} yet well
understood.  We will illustrate some of these issues by calculating
the effects of thermal magnetic noise on spin relaxation in a typical
MRFM environment.

We model the magnetic tip as a sphere with radius $r = 1\,\mu\text{m}$
and uniform magnetization $\mu_{0}M= 1\ \text{T}$.  The electron
magnetic moment is located at a separation $d=50\ \text{nm}$ from the
tip surface.  At this distance the polarizing field is $B_{0} = 2
\mu_{0}M r^{3}/(3(r+d)^{3}) = 0.58\ \text{T}$, and the spin precession
frequency is $\gamma B_{0}/(2\pi) = 16.1\ \text{GHz}$, where 
$\gamma/(2\pi) =
2.8\ \text{MHz}/T$ is the gyromagnetic ration of the electron.  We
take the ambient temperature to be $T=4\ \text{K}$.  We note that
$d\ll r$ and we therefore---roughly---model the tip as a slab
of thickness $t=d$.  Finally, the direction of the polarizing field is
taken to be parallel to the vector normal to the tip suface ($\theta =
0$).

With the parameters $\{d,t,T,\gamma,\omega_{0},\theta\}$ now specified
(see Section~\ref{Section: Device parameters} for further discussion),
the electron spin relaxation rates $1/T_{1}(\sigma)$ and
$1/T_{2}(\sigma)$ are predicted as functions of the tip conductivity
$\sigma$ by expressions (\ref{eq:G1expanded}--\ref{eq:G2expanded}),
with results as shown in Fig.~\ref{fig:Relaxation}.  For completeness,
relaxation times for proton magnetic moments at the same distance from
the tip are also shown.

\subsubsection{Design lessons for MRFM} It is clear that tip
conductivity is a key engineering parameter in MRFM. For example, if
the tip conductivity is as great as that of pure elemental iron
($\sim4\times 10^{9}\ \Omega\text{-m}$ at $4\ \text{K}$) or nickel
($\sim1.7\times 10^{10}\ \Omega\text{-m}$ at $4\ \text{K}$), the predicted
relaxation rate is $1/T_{1} \gtrsim 10^{2}\ \text{s}^{-1}$.  Such
rates are much too rapid to permit coherent spin imaging.

Real-world MRFM tips typically are composed not of pure elements, but
of magnetically ``hard'' alloys such as samarium-cobalt or
neodymium-iron-boron.  These alloys are commerically prepared as
powders, without much regard for minimizing lattice defects: the
resulting conductivity is very likely to be substantially reduced
relative to the values quoted above for pure elements.  There is not
much data in the literature relating to the electrical conductivity of
ferromagnetic particles at cryogenic temperature.  Misra \emph{et
al.}\ \cite{Misra:99} have measured the conductivity of sputtered
Cu/Cr multilayered films to be $\sim2.5\times 10^{-7}\
\Omega\text{-m}$ in the zero-temperature limit.  If we take this value
to be representative of micron-scale ferromagnetic tips at cryogenic
temperatures---in the absence of better data---the predicted
relaxation rate is $1/T_{1} \sim 10\ \text{s}^{-1}$.  This is still an
uncomfortably rapid relaxation rate.

If these predictions are correct, the MRFM community has at least five
design options: (1) reduce the temperature, (2) reduce the tip
conductivity (\emph{e.g.}, by employing an insulating magnetic garnet
tip), (3) employ superconducting tips (\emph{e.g.}, generating
gradients via trapped vortices), (4) detect electron moments
incoherently, or (5) focus on coherent proton detection, for which the
predicted relaxation times are much longer.

We see how vital it is to achieve a thorough understanding of all the
mechanisms that influence spin relaxation.  It is sobering, therefore,
to reflect upon some of the thermal reservoirs that we have \emph{not}
mentioned, which surely are present in ferromagnetic tips: spin waves,
thermally excited domain wall motions, and unpaired electron spins in
passivating oxide layers, to name three.  As discussed in
Section~\ref{Section: Conclusions}, much work remains to be done.

%
\subsection{Quantum computing}
\label{Section: Quantum computing}

Kane \cite{Kane:98} has described a design for a solid-state quantum
computer in which the qubits are the electron and nuclear spin quantum
numbers of \thirtyoneP donor sites in a silicon lattice.  Now we will
consider the effects of thermal magnetic noise within Kane's device.

In Kane's design, the \thirtyoneP donor sites carry both nuclear and
electron quantum numbers, which are coupled by a Hamiltonian
$H_{\text{int}}$ of the form
\begin{equation}
	H_{\text{int}} = 
	\gamma_{\text{e}} \lb{B}_{0} \gsb\cdot \lb{s}_{\text{e}}
	+ \gamma_{\text{n}} \lb{B}_{0} \gsb\cdot \lb{s}_{\text{n}}
	+ 4\hbar^{-2}\!A\,\lb{s}_{\text{e}} \gsb\cdot \lb{s}_{\text{n}}
\end{equation}
Here $\lb{s}_{\text{e}}$ and $\lb{s}_{\text{n}}$ are the electron and
\thirtyoneP spin operators, $\gamma_{\text{e}}$ and
$\gamma_{\text{n}}$ are the gyromagnetic ratios (in a convention where
both ratios are positive), and the hyperfine coupling $A$
has the value $\hbar A/(2\pi) = 29\ \text{MHz}$.

The computational action occurs between the two lowest energy levels
of this system, and the presence of the hyperfine coupling modifies
these lowest-energy levels in two respects.  First, the energy
separation of the two levels depends on $A$ via an equation given by
Kane as
\begin{equation}
	\hbar \omega_{0} = \gamma_{\text{n}} B_{0} + 2 A  + \lcal{O}(A^{2})
\end{equation}
Second, the two lowest energy levels of the system couple to external
time-dependent B-fields $\lb{B}_{\text{ext}}(t)$ via an effective
Hamiltonian $H_{\text{eff}}$ which a straightforward perturbative
calculation shows to be
\begin{equation}
	H_{\text{eff}} = \frac{\hbar \gamma_{\text{n}}}{2}\,
	\lb{B}_{\text{ext}}(t) \gsb\cdot \lb{K} \gsb\cdot \gsb{\sigma}
\end{equation}
where $\gsb{\sigma} \equiv\{\sigma_{x},\sigma_{y},\sigma_{z}\}$ is a
vector of Pauli matrices acting on the two lowest-energy states, and
$\lb{K}$ is a dimensionless matrix given by
\begin{equation}
	  \lb{K} = \lbhat{b}\gsb\otimes\lbhat{b} + \left(1+\frac{2
	  A}{\gamma_{\text{n}}B_{0}}\right) \big(\lb{I}-
	  \lbhat{b}\gsb\otimes\lbhat{b}\big) + \lcal{O}(A^{2})
\end{equation}
where $\lbhat{b}$ is a unit vector in the direction of the applied
B-field.  We observe that the hyperfine coupling $A$ acts to increase
the coupling of the two lowest-energy states to thermal magnetic
noise.  This increased coupling is most conveniently taken into
account by the replacement in \eqref{eq:RotatingSpectralDensity} of
the laboratory-frame spectral density $\lb{S}_{\lb{B}}$ by an
effective spectral density $\lb{S}_{\lb{B}}^{\text{eff}}$
\begin{equation}
\lb{S}_{\lb{B}} \rightarrow\lb{S}_{\lb{B}}^{\text{eff}}
\equiv \lb{K}\gsb\cdot\lb{S}_{\lb{B}}\gsb\cdot\lb{K}
\end{equation}
Subsequent calculations of relaxation rates are unaltered.  In Kane's
design, the renormalized coupling $\lb{K}$ increases the predicted
relaxation rates by a factor of order
$\left(1+(2A)/(\gamma_{\text{n}}B_{0})\right)^{2}\sim 7.2\,$.

Per Kane's design \cite{Kane:98}, we specify a polarizing field $B_{0}
= 2\ \text{T}$ and a temperature $T = 100\ \text{mK}$.  We model
Kane's A-gates and J-gates as metallic pads of thickness $t=5\
\text{nm}$ at a distance $d=10\ \text{nm}$ from the \thirtyoneP
centers.  The direction of the polarizing field is taken to be
parallel to the vector normal to the pad surface ($\theta = 0$).  The
electron-spin and nuclear-spin relaxation rates $1/T_{1}$ and
$1/T_{2}$ are then predicted as functions of the pad conductivity
$\sigma$ by expressions (\ref{eq:G1expanded}--\ref{eq:G2expanded}),
with results as shown in Fig.~\ref{fig:Relaxation}. 

\subsubsection{Design lessons for solid-state quantum computing}
Kane's article optimistically quotes the magnetic resonance literature
to the effect 
\begin{quote}
at temperatures $T\sim 1.5\ \text{K}$ the electron
spin relaxation time is thousands of seconds, and [\ldots] at
millikelvin temperatures the phonon limited \thirtyoneP relaxation
time is of the order of $10^{18}$ seconds, making this system ideal
for quantum computation.
\end{quote}
We see that thermal magnetic noise originating in the A- and J-gates
is likely to substantially increase these relaxation rates. 
Fortunately, to the extent that calculation occurs only between the
two lowest-energy states, the predicted rapid relaxation of the
higher-energy electron states may not pose a problem.  And it is
gratifying that the predicted \thirtyoneP relaxation rates remain
reasonably small even when the amplifying effects of the hyperfine
coupling are taken into account.

These predictions should be regarded very cautiously, however.  As we
discuss at greater length in the following section, even the notion of
conductivity becomes suspect at these length scales and temperatures. 
In our view, the main design lesson is that all potential noise
sources must be treated seriously.

\section{Conclusions}
\label{Section: Conclusions}

We have developed a general theory of thermal magnetic fluctuations
near the surface of conductive and/or magnetically permeable slabs. 
We have applied the resulting closed-form expression for the magnetic
spectral density (Section~\ref{sec:ThermalMagneticSpectralDensities})
to practical problems in atomic physics experiments, magnetic
resonance force microscopy, and solid-state quantum computing
(Sections~\ref{Section: Atom and ion traps}--\ref{Section: Quantum
computing}).  The theory predicts magnetic noise levels that are large
enough to require substantial care to minimize their effects.

More generally, our formalism indicates that if the dissipative kernel
of a linearly damped system is known, then the consequences for
fluctuation, dissipation, entanglement, and renormalization are fully
determined, via fluctuation-dissipation theorems {\large
(}eqs.~\eqref{eq:GammaNoise} and \eqref{eq:SfFor Oscillators}{\large
)}, fluctuation-dissipa- tion-entanglement theorems {\large
(}eqs.~\eqref{eq:fde theorem} and \eqref{eq:harmonicFDE}{\large )},
and renormalization-dissipation relations {\large
(}eqs.~\eqref{eq:spinRD} and \eqref{eq:harmonicRD}{\large )}.

We postulate that these
fluctuation-dissipation-entanglement-renormaliz-ation relations are
\emph{universal}, in the sense that they apply to \emph{all} thermal
reservoirs, without regard for the internal structure of the
reservoir, with the sole restriction that the coupling to the
reservoir must be weak enough to be linearly dissipative.

A major limitation of our article is that we derived our results in
the context of a specific model of thermal reservoirs---the
independent oscillator (IO) model---and hence we have \emph{not}
proved universality.  Ford, Lewis, and O'Connell's pioneering article
on independent oscillator models~\cite{Ford:88} proves universality in
the restricted sense that these models are shown to describe all forms
of fluctuation and dissipation that are causal and linearly
dissipative.  An important milestone for further research would be to
prove universality in an expanded context which included entanglement
and renormalization in addition to fluctuation and dissipation.

There are also substantial reasons to expect that measurement
processes might also be subsumed under a suitably expanded notion of
universality.  We reason as follows.  Quantum-coherent technologies
necessarily include some means for reading-out binary data; in quantum
computing the data are qubits, in MRFM experiments the data are the
presence or absense of a spin at a given atomic coordinate.  In most
cases measurement is a continuous interferometric process, achieved
\emph{e.g.}\ via weak interactions of a cantilever with laser-supplied
photons.\footnote{We remark that the detailed quantum dynamics of the
interferometer-cantilever-spin system remains to be worked out.} From
an engineering point of view, the measurement photons comprise a
``good'' reservoir---effectively at zero temperature---whose finely
tuned interactions compete with many different ``bad'' thermal
reservoirs that contribute only noise.  The quantum system seeks to achieve
equilibrium with the ``good'' and ``bad'' reservoirs simultaneously;
the resulting quantum dynamics are traditionally described by master
equations of the Fokker-Planck type.  A unified formalism of this
sort, treating measurement and noise processes on a common footing,
would offer many practical advantages in quantum-coherent engineering.

In carrying through these calculations, we have come to appreciate an
emerging common ground---the technical challenge of preserving and
manipulating quantum coherence---that is shared by magnetic resonance
force microscopy and quantum computation.  Single-spin MRFM imaging
devices (if they can be demonstrated) would constitute first-generation
solid-state quantum computers, in which the individual qubits carried
binary information about the presence or absence of a spin spin at a
specified atomic coordinate.

Arguably the most daunting challenge in developing quantum-coherent
technologies is achieving simultaneous engineering control of a
panoply of thermal reservoirs:
\begin{itemize}
	\item phonons and surface vibrational modes 
	\item ferro- and ferrimagnetic spin waves and domain walls
	\item conduction band excitations in metals and semiconductors
	\item paramagnetic and nuclear spins 
	\item mobile molecules and charges on surfaces
	\item lattice defects and impurities
	\item super-conducting condensates admixed with 
	normal-phase conductors
\end{itemize}
This list includes most of the excitations that are commonly studied
by condensed matter physicists.  We remark it is not enough to achieve
good engineering control of some or even most of these noise sources;
\emph{all} must be controlled simultaneously, as even one
poorly-controlled noise source is enough to destroy quantum coherence. 

The dissipative kernels of these excitations are poorly understood at
the ultra-low temperatures and mesoscopic length scales that are
characteristic of quantum-coherent technologies.  For example, even
Maxwell's equations in conducting media---an exceedingly well-studied
area of physics---will require re-interpretation due to our
still-limited understanding of mesoscopic conductivity
\cite{Landauer:96, Landauer:99}.

In view of these difficulties, Landauer has modestly proposed
\cite{Landauer:99b} that the following disclaimer be appended to
all quantum computing proposals:
\begin{quote}
	\label{quote: Landauer}
   This proposal, like all proposals for quantum computation, relies
   upon speculative technology, does not in its current form take into
   account all possible sources of noise, unreliability and
   manufacturing error, and probably will not work.
\end{quote}
Our article may be read as a response to Landauer's concerns: we have
analyzed thermal magnetic noise as a ``possible source of noise and
unreliability'' in MRFM imaging and in Kane's proposal for a quantum 
computer.  More ambitiously, we have done so in the context of
a general formalism which might be applied across a broad range of
thermal reservoirs.  In doing so, we hope to have contributed to the
emerging task of providing solid and well-organized foundations for
the development of quantum-coherent technologies.

%% file: MRFM_Magnetic_Part_02.tex
\section{Fluctuation and dissipation}
\label{Appendix: Fluctuation and dissipation}

Many readers will be acquainted with the fluctuation-dissipation
theorem as it applies to voltage noise---known as Johnson noise---in
resistive circuit elements \cite{Landau:80Thermal}.  The theorem
can be briefly stated as follows: if a net charge $q(t)=q e^{i\omega
t}$ is passed through a frequency-dependent impedance $Z(\omega)$,
such that the current $i(t)= \partial q(t)/\partial t = i\omega q
e^{i\omega t}$, and the power $P(\omega)$ dissipated in the impedance
is $P(\omega) =\text{Re}(Z(\omega))\,q^{2}\omega^{2}$, then the
spectrum of thermal voltage fluctuations is given in terms of the
dissipative impedance $\text{Re}(Z)$ by \eqref{eq:RNoise}.

The general fluctuation-dissipation theorem allows us to deduce, by an 
exact analogy, that if a time-dependent magnetic moment $\lb{m}(t) = 
\lb{m}\,e^{i\omega t}$ is adjacent to a conductive slab, such that the 
power dissipated in the slab is 
$P=\lb{m}\gsb\cdot\gsb\Gamma(\omega)\gsb\cdot\lb{m}\,\omega^{2}$, 
then the spectrum of the magnetic fluctuations is given in terms of 
the magnetic dissipation matrix $\gsb\Gamma(\omega)$ via 
\eqref{eq:GammaNoise}.

Discussions which justify the above line of reasoning are presented in
many graduate-level textbooks \cite{Landau:80aFDtheorem,Balescu:75},
but these discussions tend to be rather abstract and lengthy.  Some
readers will prefer the following physical derivation, which is
completely general and rigorous.

We will deduce the spectrum of magnetic fluctuations from the
well-known spectrum of thermal voltage fluctuations
\cite{Landau:80Thermal} (Johnson noise).

Let $A$ be the area of the pickup loop in Fig.~\ref{fig:SkinCoil}, and 
let \lbhat{m} be the unit vector normal to the loop.  We let $I$ and 
$\lb{m}$ be the real coefficients of the loop current $I(t) = I e^{i\omega 
t}$ and loop magnetic moment $\lb{m}(t) = \lb{m} e^{i\omega t}$,
such that $\lb{m} = A I\lbhat{m}$.  We wish to 
compute the spectral density $ 
\lbhat{m}\gsb\cdot\lb{S}_{\lb{B}}\gsb\cdot\lbhat{m}$ of the magnetic 
field traversing the loop.  We reason as follows:\footnote{We write $2 
k_{\text{B}}T$ in place of $\hbar 
\omega\coth(\hbar\omega/(2k_{\text{B}}T))$ purely for economy of 
notation.}
\[
\lbhat{m}\gsb\cdot\lb{S}_{\lb{B}}\gsb\cdot\lbhat{m}\ 
\begin{array}[t]{ll}
	=\ S_{\Phi}/A^{2},&
	\text{magnetic flux $\Phi=A\,\lbhat{m}\gsb\cdot\lb{B}$};\\[1ex]
	=\ S_{V}/(\omega A)^{2},&
		\text{Faraday's law of Induction;}\\[1ex]
	=\ \text{Re}(Z)/(\omega A)^{2}\ 2 k_{\text{B}}T,&
		\text{thermal voltage noise per \eqref{eq:RNoise};}\\[1ex]
	=\ P/(\omega I A)^{2}\ 2 k_{\text{B}}T,&
		\text{because $P=\text{Re}(Z)I^{2}$;}\\[1ex]
	=\ \lb{m}\gsb\cdot\gsb\Gamma\gsb\cdot\lb{m}/(I A)^{2}\ 2 k_{\text{B}}T,&
		\text{definition of $\gsb\Gamma$ in terms of $P$;}\\[1ex]
	=\ \lbhat{m}\gsb\cdot\gsb\Gamma\gsb\cdot\lbhat{m}\ 2 k_{\text{B}}T,&
		\text{substitute $\lb{m}=IA\,\lbhat{m}$.}
\end{array}
\]
Since the result holds for arbitrary loop orientation $\lbhat{m}$, and
both $\lb{S}_{\lb{B}}$ and $\gsb{\Gamma}$ are symmetric matrices, it
must be that $\lb{S}_{\lb{B}}(\omega) = \gsb\Gamma(\omega)\,2 k_{\text{B}}T$, 
\emph{QED.}
\subsection{Spatial correlations and gradients}
\label{Section: Spatial correlations and gradients}
This reasoning is readily extended to encompass spatial 
correlations and gradients.  Let $n$ magnetic moments 
$[\lb{m}]\equiv\{\lb{m}_{1}, \lb{m}_{2}, \ldots,\lb{m}_{n}\}$ be 
located at coordinates $\{\lb{x}_{1}, \lb{x}_{2},\ldots, 
\lb{x}_{n}\}$.  Assuming a time dependence $\propto e^{i\omega t}$, 
the dissipated power is calculable from the geometry and conductivity 
of the system---at least in principle---and will be of the general 
form $P = [\lb{m}]\gsb\cdot\left[\gsb\Gamma\right] 
\gsb\cdot[\lb{m}]\,\omega^{2}$, where $\left[\gsb\Gamma\right]$ is the 
obvious $n$--point generalization of $\gsb{\Gamma}$.  Then the 
$n$--point generalization of \eqref{eq:GammaNoise} is simply
\begin{equation}
\lb{S}_{[\lb{B}]}= \left[\gsb\Gamma\right]\,
	\hbar\omega \coth\left(\dfrac{\hbar\omega}{2k_{\text{B}}T}\right)
\end{equation}
In this fashion all questions involving $n$--point spectral densities 
of magnetic noise can be reduced to the calculation of 
$\left[\gsb\Gamma\right]$.  Although we will not do so in this 
article, the spectral density of magnetic gradients can be determined 
by considering the limiting case in which two coordinates approach 
each other.

\subsection{Classical and quantum calculations}
Students in particular may be troubled that our calculation of the
dissipation tensor $\gsb\Gamma(\omega)$ will be entirely classical,
via Maxwell's equations.  After all, doesn't the title of this article
promise ``the classical \emph{and quantum} theory of thermal magnetic
noise''?  The quantum part of the promise will be fullfilled in
Appendix~\ref{Appendix: Quantum decoherence}, to follow, in which we
present a rigorous quantum theory of a spin coupled to a thermal
reservoir, and show that the classical dissipative kernel determines 
the quantum dissipation also.

\subsection{Solving the Maxwell equations}
\label{sec:SolvingTheMaxwellEquations}
To calculate power dissipation within the slab, and thus determine 
$\gsb\Gamma(\omega)$, we seek solutions of Maxwell's equations in 
their standard form for time dependence 
$\propto e^{i\omega t}$ \cite{Reitz:67Maxwell}:
\begin{align}
\label{eq:Maxwell}
\gsb\nabla \gsb\times\lb{E} &= -i \omega \lb{B}&
 \gsb\nabla\gsb\cdot\lb{D} &=\rho_{\text{c}} 
 \\ \notag
 \gsb\nabla\gsb\times\lb{H}& = \lb{j}+i\omega\lb{D}  & \gsb\nabla\gsb\cdot\lb{B} &=0 
 \\
\intertext{together with the constitutive and divergence equations for 
 the current}
\lb{j} & = (\sigma\lb{E} + 
 \lb{j}^{\text{s}}) & \gsb\nabla\gsb\cdot\lb{j} &=-i\omega \rho_{\text{c}}
\end{align}
Here \lb{B} is the magnetic field, $\lb{H}=\lb{B}/\mu$ is the magnetic 
intensity, $\mu$ is the magnetic permeability, \lb{E} is the electric 
field, $\lb{D}=\epsilon\lb{E}$ is the electric displacement, $\lb{j}$ 
is the total current, $\epsilon$ is the electric permittivity, 
$\rho_{\text{c}}$ is the charge density, $\sigma$ is the conductivity, 
and $\lb{j}^{\text{s}}$ is the externally-applied dipole source current.

\subsubsection{The scalar Maxwell equations} 
\label{sec:The scalar Maxwell equations}

We begin our solution of the Maxwell equations with the help of a 
simplifying \emph{ansatz}: that the irrotational part of $\lb{E}$ and 
$\lb{D}$ vanishes, \emph{i.e.}, the charge density $\rho_{\text{c}}$ 
is zero within the slabs and at their interfaces.  Furthermore, we 
will assume that the displacement current 
$i\omega\lb{D}=i\omega\epsilon\lb{E}$ is negligibly small compared to 
the conduction current $\sigma\lb{E}$.  This approximation is quite 
accurate for all the cases we will consider, \emph{e.g.}, for room 
temperature copper at one GHz we have $\omega\epsilon/\sigma \sim 
10^{-9}$.\footnote{We remark that the irrotational \emph{ansatz} and 
the neglect of displacement currents must be embraced simultaneously, 
because individually the \emph{ansatz} and the approximation yield a 
mathematically inconsistent set of Maxwell equations.  See, 
\emph{e.g.}, \cite{Reitz:67Maxwell} for a discussion.}

It is natural to write a divergence-free \lb{E} field as a curl; we 
write \lb{E} as
\begin{subequations}
\begin{equation}
\label{eq:Econstitutive}
\lb{E} = -i\omega\,\gsb\nabla\gsb\times (\psi\lbhat{n}) 
\end{equation}
where $\psi$ is a scalar vorticity, and we recall that \lbhat{n} is 
the (constant) unit vector normal to the slab surfaces.  Per our 
\emph{ansatz}, \eqref{eq:Econstitutive} guarantees that 
$\gsb\nabla\gsb\cdot\lb{D}=0$ everywhere within a slab (boundary 
conditions will be considered shortly).  In terms of $\psi$, the 
Maxwell equation $\gsb\nabla \gsb\times\lb{E} = -i \omega \lb{B}$ 
becomes
\begin{equation}
\label{eq:Bconstitutive}
\lb{B} = \lbhat{n}\,(\nabla^{2}\psi) -
	\gsb\nabla(\lbhat{n}\gsb\cdot\gsb\nabla\psi)
\end{equation}
which, furthermore, guarantees $\gsb\nabla\gsb\cdot\lb{B} = 0$.  
The sole remaining Maxwell equation $\gsb\nabla\gsb\times\lb{H}= 
\sigma\lb{E} + \lb{j}^{\text{s}}$ is then equivalent to the scalar wave 
equation
\begin{equation}
\label{eq:scalarWaveEquation}
	\nabla^{2}\psi = i\mu \sigma \psi + j^{\text{s}}
\end{equation}
where $j^{\text{s}}$ is a scalar dipole source whose nature will be 
specificed shortly.

It remains only to enforce the usual boundary conditions of 
electrodynamics: continuity of the normal components of $\lb{B}$ and 
$\lb{D}$ and the transverse components of $\lb{E}$ and $\lb{H}$.  
These are equivalent to the following boundary conditions on $\psi$ at 
the juncture of two slabs $i$ and $j$
\begin{align}
\label{eq:boundary}
	\psi_{i} &= \psi_{j}\\
	(\lbhat{n}\gsb\cdot\gsb\nabla \psi_{i})/\mu_{i}  &=
	(\lbhat{n}\gsb\cdot\gsb\nabla \psi_{j})/\mu_{j}
\end{align}
\end{subequations}
The problem is now reduced to the solution of the scalar wave 
equation \eqref{eq:scalarWaveEquation}, subject to the 
between-slab boundary conditions \eqref{eq:boundary}.  The \lb{E} 
and \lb{B} fields are given by \eqref{eq:Econstitutive} and 
\eqref{eq:Bconstitutive}, which now are purely constitutive 
equations.




\subsubsection{The scalar dipole source} 
\label{sec:TheScalarDipoleSource} 
For a (vector) magnetic 
dipole \lb{m} located within a region of zero conductivity, it is 
convenient to specify the scalar source $j^{\text{s}}$ implicitly by 
$j^{\text{s}}=\nabla^{2}\psi^{\text{s}}$, with the source field 
$\psi^{\text{s}}$ given by
\begin{equation}
\label{eq:psiequation}
\psi^{\text{s}}(\lb{x}) = \frac{\mu}{4 \pi}\, 
\left(\frac{\lbhat{n}}{|\lb{x}|} + 
\frac{\lbhat{x}\cdot(\lb{I}-\lbhat{n}\otimes\lbhat{n})} 
{|\lb{x}|+\lb{x}\cdot\lbhat{n}} \right)\cdot\lb{m}
\end{equation}
Inserting $\psi^{\text{s}}$ in the constitutive equation
\eqref{eq:Bconstitutive} yields for \lb{B}
\begin{equation}
\label{eq:Bequation}
\lb{B}(\lb{x}) = 
\frac{\mu}{4 \pi |\lb{x}|^{3}}\,
(3\,\lbhat{x}\otimes\lbhat{x}-\lb{I})\cdot\lb{m}
\end{equation}
which (by construction) is the dipole B-field generated by a magnetic 
moment \lb{m}.  This justifies our specification of $\psi^{\text{s}}$ as 
the source field.  The possibility of writing a general dipole field 
in this form is the key step leading to the success of our scalar 
\emph{ansatz}.

As a technical point, the vector potential 
$\psi^{\text{s}}(\lb{x})\lbhat{n}$ describes an E-field which is 
singular along a line extending from $\lb{x}=0$ to infinity in the 
$-\lbhat{n}$ direction (away from the slab).  As we will show in 
Section~\ref{sec:RepairingTheElectricField}, the singular portion of 
the E-field is irrotational, and it can be written as the gradient of a 
harmonic potential which disappears when we introduce a finite charge 
density on the surface of the slab.  The dissipated power within the 
slab is not appreciably affected by the currents required to create 
this surface charge, so for the time being the singularity can simply 
be ignored.

\subsubsection{Solving the scalar Maxwell equations}
The scalar equation \eqref{eq:scalarWaveEquation} for $\psi$ is most 
easily solved in cylindrical coordinates $\{r,\varphi,z\}$.  It is 
further convenient to work with the Bessel transform 
$\stilde\psi_{m}(\rho,z)$ defined by\footnote{The Bessel functions 
$J_{m}$ satisfy an orthogonality relation \[ 
\int_{0}^{\infty}\!\!r\,dr\ J_{m}(\rho r)J_{\sprime{m}}(\sprime\rho r) 
= \delta_{m\sprime{m}}\,\delta(\rho-\sprime{\rho})/\rho\]}
\begin{equation}
\label{eq:BesselJ}
	\stilde\psi_{m}(\rho,z) = 
	\frac{1}{2\pi}\int_{0}^{2\pi}\!d\varphi
	\int_{0}^{\infty}\!r\,dr\,\,e^{-im\varphi}
	J_{m}(\rho r)\,\psi(r,\varphi,z)
\end{equation}
whose inverse is
\begin{equation}
\label{eq:inverse}
	\psi(r,\varphi,z) = \sum_{m=-\infty}^{m=\infty}e^{im\varphi}\int_{0}^{\infty}\!d\rho\,\rho\,
	J_{m}(\rho r) \tilde\psi_{m}(\rho,z)
\end{equation}
Here $J_{m}$ is a Bessel function of integer argument.  The scalar 
wave equation for $\stilde\psi_{m}(\rho,z)$ is
\begin{equation}
\frac{\partial^{2}}{\partial z^{2}} \stilde\psi_{m}(\rho,z) = 
i\omega\mu\sigma\stilde\psi_{m}(\rho,z) + \tilde{j}_{m}^{\text{s}}(\rho,z)
\end{equation}
whose general homogenous solution is
\begin{equation}
\label{eq:BesselCoefficients}
\stilde\psi_{m}(\rho,z) = \tilde\psi^{+}_{m}(\rho)e^{kz}+\tilde\psi^{-}_{m}(\rho)e^{-kz}
\end{equation} 
where we have introduced the complex wavenumber 
$k\equiv\sqrt{i\omega\mu\sigma+\rho^{2}}$.

Our next task is to explicitly solve for the Bessel coefficients 
$\tilde\psi^{\rm}_{m}(\rho)$.  We divide space into three regions 
\{I,II,III\}, where Region~I is the nonconductive region that contains 
the dipole source, Region~II is the conductive slab of thickness $t$ 
beginning at a distance $d$ from the source, and Region~III is the 
nonconducting region on the opposite side of the slab (see 
Fig.~\ref{fig:Regions}).

Within Regions~I and III we assume vanishing conductivity $\sigma=0$ 
and a purely real permeability $\mu_{0}$.  In contrast, within 
Region~II we allow both $\sigma$ and $\mu$ to assume complex values.  
Thus energy dissipation can occur only within Region~II.

Requiring that $\stilde\psi_{m}(\rho,z)\to0$ as $|z|\to\infty$ leads to the 
general solution:
\begin{equation}
\label{eq:LinearEquations}
\stilde\psi_{m}(\rho,z)=\left\{
\begin{array}{ccl}
\stilde\psi^{\text{I}{+}}_{m}(\rho)e^{\rho z}+\stilde\psi^{\text{s}}_{m}(\rho,z)& 
& 
\text{Region I (source);} \\
\parbox[c]{12em}{\dotfill}&
|&
{\scriptstyle\ z=d}\\
\stilde\psi^{\text{II}{+}}_{m}(\rho) e^{kz}+\stilde\psi^{\text{II}{-}}_{m}(\rho) e^{-kz}&
\!\!+z\!\!&\text{Region II (slab);} \\
\parbox[c]{12em}{\dotfill}&
\downarrow&
{\scriptstyle\ z=d+t}\\
\stilde\psi^{\text{III}{-}}_{m}(\rho) e^{-\rho z}&
&
\text{Region III (far side).}
\end{array}
\right.
\end{equation}
For illustrations of Regions I, II, and III, see 
Figs.~\ref{fig:Regions}--\ref{fig:SkinCoil}.  The source term 
$\stilde\psi^{\text{s}}_{m}(\rho,z)$ in Region~I is found by 
substituting \eqref{eq:psiequation} in \eqref{eq:BesselJ}:
\begin{align}
\label{eq:oneHalf}
\stilde\psi^{\text{s}}_{0}(\rho,z) &= \frac{\mu_{0}}{4\pi\rho}\,e^{-\rho 
|z|} \lbhat{n}\gsb\cdot\lb{m} \\
\stilde\psi^{\text{s}}_{\pm1}(\rho,z) &= \frac{\mu_{0}}{8\pi\rho}\,e^{-\rho 
|z|}\,(\lbhat{x}\pm i\lbhat{y})\gsb\cdot\lb{m}
\end{align}
where we have introduced Cartesian basis vectors 
$\lbhat{x}\gsb\times\lbhat{y}=\lbhat{n}$.  Note the singularity 
at $z=0$; this reflects our assumption of a point dipole source.  
Section~\ref{sec:RepairingTheElectricField} derives explicitly 
nonsingular expressions for $\lb{B}$ and $\lb{E}$ originating from 
finite-sized current sources, however the present dipole result 
is adequate for our main purpose of estimating power dissipation in the 
slab.
 
With the source term now specified, the boundary equations 
\eqref{eq:boundary} for the Bessel coefficients 
$\{\stilde\psi^{\text{I}{+}}_{0}, \stilde\psi^{\text{II}{+}}_{0}, \stilde\psi^{\text{II}{-}}_{0}, 
\stilde\psi^{\text{III}{-}}_{0}\}$ can be readily solved.  We obtain for the 
$m=0$ coefficients:
\begin{multline}
\label{eq:BesselCoefficientResults}
  \left[
  \begin{matrix}
    \stilde\psi^{\text{I}{+}}_{0}\\[0.5ex]
    \stilde\psi^{\text{II}{+}}_{0}\\[0.5ex]
    \stilde\psi^{\text{II}{-}}_{0}\\[0.5ex]
    \stilde\psi^{\text{III}{-}}_{0}
  \end{matrix}
  \right] = \frac{\mu_{0}}{4\pi}\ 
   \left[
  \begin{matrix}
    e^{-2\rho d}(e^{2kt}-1)(K^{2}\rho^{2}-k^{2})\\[0.5ex]
    -2e^{-(\rho+k)d}(K\rho-k)K\rho\\[0.5ex]
    2e^{2kt-(\rho-k)d}(K\rho+k)K\rho\\[0.5ex]
    4e^{(\rho+k)} k K\rho
  \end{matrix}
  \right]
  \times\\
    \frac{\lbhat{n}\gsb\cdot\lb{m}}
      {\rho\big((K^{2}\rho^{2}+
         k^{2})(e^{2kt}-1)+2K\rho k(e^{2kt}+1)\big)}
\end{multline}
where $K\equiv\mu/\mu_{0}$ is the relative permeability of the slab.  
The $m=\pm1$ coefficients are directly proportional to the $m=0$ 
coefficients 
\begin{align}
	\left[
	\begin{matrix}
	  \stilde\psi^{\text{I}{+}}_{\pm1}\\
      \vdots
	\end{matrix}
	\right] &= 	
	\frac{(\lbhat{x}\pm i\lbhat{y})\gsb\cdot\lb{m}}
	{2(\lbhat{n}\gsb\cdot\lb{m})}	
	\left[
	\begin{matrix}
	  \stilde\psi^{\text{I}{+}}_{0}\\
      \vdots
	\end{matrix}
	\right]
\end{align}
When we calculate magnetic spectral densities, this simple proportionality will ensure that
$\gsb\Gamma\propto(\lb{I}+\lbhat{n}\gsb\otimes\lbhat{n})$ at all 
frequencies and length scales.

\subsubsection{Energy dissipation}  Our next task is to calculate the
power $P$ dissipated within Region~II.  From classical electrodynamics
we have
\begin{align}
P &= \int_{V}\!\!dV\ \left(\lb{j}\gsb\cdot\lb{E} + 
\lb{H}\gsb\cdot\frac{\partial\lb{B}}{\partial t}\right)\\
\intertext{which with the help of the Bessel orthogonality relations 
becomes} P &= 2\pi\!\!\!
\sum_{m=-\infty}^{\infty}\int_{0}^{\infty}\!\!\!\rho\,d\rho 
\int_{d}^{d+t}\!\!\!\!dz\ \left(\rho^{2}\omega^{2}\text{Re}(\sigma) + 
\rho^{4}\omega\frac{\text{Im}(\mu)}{|\mu|^{2}}\right)\ 
|\psi_{m}(\rho,z)|^{2}
\end{align}
Carrying through the $z$-integration, we find the scalar magnetic 
dissipation coefficient $\Gamma(\omega)$, as implicitly defined by $P 
\equiv \omega^{2}\Gamma(\omega)\,\big(\lb{m}\gsb\cdot(\lb{I} + 
\lbhat{n}\gsb\otimes\lbhat{n})\gsb\cdot\lb{m}\big)$\,:
\begin{multline}
\label{eq:integrand}
\Gamma(\omega) = 
\frac{\mu_{0}^{2}}{4\pi}\int_{0}^{\infty}\!\!d\rho\ 
\left(\rho^{3}\text{Re}(\sigma) + 
 \frac{\rho^{5}\text{Im}(\mu)}{\omega |\mu|^{2}}\right)\ |K|^{2}e^{-2\rho d}\times\\
 \frac{\text{Re}\left[ik^{\ast}(e^{2kt}-1)(e^{2k^{\ast}t}+1)
    (K^{2}\rho^{2}-k^{2})/(\omega\mu\sigma) +
    \rho K |e^{2kt}-1|^{2}\right]}
 {\big|(K^{2}\rho^{2}+k^{2})(e^{2kt}-1)+2K\rho k(e^{2kt}+1)\big|^{2}}
\end{multline}
where we recall that $k\equiv\sqrt{i\omega\mu\sigma+\rho^{2}}$ and 
$K\equiv\mu/\mu_{0}$.  In simplifying this result we have kept only 
the leading and next-to-leading terms in $\text{Im}(\mu)$, 
\emph{i.e.}, we have assumed $K$ is predominantly real.

The case of thermal magnetic fluctuations measured at a point midway 
between two identical conducting slabs can be solved by exactly 
similar methods.  It is only necessary to replace the Region~I 
homogenous term $\stilde\psi^{\text{I}{+}}_{m}(\rho)e^{\rho z}$ in 
\eqref{eq:LinearEquations} with the symmetrized term 
$\stilde\psi^{\text{I}{+}}_{m}(\rho)\cosh(\rho z)$.  The resulting magnetic 
dissipation coefficient is
\begin{multline}
\sprime\Gamma(\omega) = 
\frac{\mu_{0}^{2}}{4\pi}\int_{0}^{\infty}\!\!d\rho\ 
\left(\rho^{3}\text{Re}(\sigma) + 
 \frac{\rho^{5}\text{Im}(\mu)}{\omega |\mu|^{2}}\right)\ |K|^{2}e^{-2\rho d}\times\\ 
 \frac{2\text{Re}\left[ik^{\ast}(e^{2kt}-1)(e^{2k^{\ast}t}+1)
    (K^{2}\rho^{2}-k^{2})/(\omega\mu\sigma) +
    \rho K |e^{2kt}-1|^{2}\right]}
 {\big|(k^{2}+K^{2}\rho^{2}+e^{-2\rho d}(k^{2}-K^{2}\rho^{2}))(e^{2kt}-1)+2K\rho k(e^{2kt}+1)\big|^{2}}
\end{multline}
Note that the double-slab integrand is just twice the single-slab 
integrand, slightly modified by an additional term $\propto e^{-2\rho 
d}$ in the denominator.  For the special case $K=1$ (the case of 
primary interest) this additional term can be shown to be non-leading 
for all values of $\{\rho, d,t,\sigma,\omega\}$.  A numerical survey 
confirms that $\sprime\Gamma(\omega)\simeq2\Gamma(\omega)$ to within 
very good accuracy, as described in 
Section~\ref{sec:ThermalMagneticSpectralDensities}.

\subsubsection{Asymptotic limits for paramagnetic and diamagnetic slabs}
The three asymptotic expressions in \eqref{eq:GammaResults} are 
obtained by evaluating \eqref{eq:integrand} for $K=1$ and 
$\text{Im}(\mu)=0$ in the following limits:
\begin{equation}
\Gamma(\omega) \simeq 
\text{Re}(\sigma)\,\frac{\mu_{0}^{2}}{4\pi}\int_{0}^{\infty}\!\!\!\!d\rho\times\!
\begin{cases}
\dfrac{(1-e^{-2\rho t})}{8 e^{2\rho d} }, & 
	\text{$|\sigma|\to0$;}\\[2.5ex]
\dfrac{\rho^{3}\lambda^{3}e^{-2\rho d}}{2\cos(\pi/4-\phi/2)}, & 
	\text{$|\sigma|\to\infty$;}\\[2.5ex]
\dfrac{\rho^{3}\lambda^{4}(t-4\rho\lambda^{2}\sin\phi)}
	{t^{2}e^{2\rho d}},  & k t\ll1\ll\rho d.
\end{cases}
\end{equation}
The integrations are readily carried out, with the results as in 
\eqref{eq:GammaResults}.  

\subsubsection{The irrotational electric field}
\label{sec:RepairingTheElectricField}
Now we have only one chore remaining: repairing the irrotational 
singularity in the E-field mentioned in 
Section~\ref{sec:TheScalarDipoleSource}, and verifying that the repair 
does not alter the energy dissipation just calculated.  In the end, we 
will obtain expressions for the E- and B-fields which are closed-form 
and explicitly finite.

In outline, the repair is accomplished by adding a purely irrotational
E-field to Region~I---thereby relaxing the \emph{ansatz} of Section~
\ref{sec:The scalar Maxwell equations} that $\lb{E}$ is purely
rotational.\footnote{We recall \emph{Helmholtz's theorem}: that any
finite vector field that vanishes at infinity can be written uniquely
as the sum of an irrotational part (the divergence of a scalar
potential) and a rotational part (the curl of a vector potential)
\cite{Morse:53HelmholtzTheorem}.} With the added E-field, Faraday's
law $\gsb\nabla \gsb\times\lb{E} = -i \omega \lb{B}$ is still
satisfied, because the added E-field is irrotational.  Because the
added field is confined to Region~I, which is nonconductive, and
because displacement currents $i\omega\lb{D}$ are negligibly small,
Ampere's law $\gsb\nabla\gsb\times\lb{H} = \sigma\lb{E} +
\lb{j}^{\text{s}}+i\omega\lb{D} \simeq \lb{j}^{\text{s}}$ is still
satisfied.  Gauss's law $\gsb\nabla\gsb\cdot\lb{D}=0$ is still
satisfied because the added E-field is the gradient of a harmonic
potential.  Physically, the added field is generated by surface charge
on the slab, and the within-slab currents necessary to sustain the
surface charge are negligibly small compared to the eddy currents
already computed.  Thus, the added E-field does not alter the magnetic
dissipation coefficient $\gsb\Gamma(\omega)$ already computed.  The
remainder of this section is devoted to proving these assertions.

We begin by noting that the E-field singularity in Region~I physically 
corresponds to the field induced by an end-to-end line of electric 
dipoles along the $-\lbhat{n}$ axis.  It is easy to verify that such a 
singularity can be cancelled by the addition of an (irrotational) 
E-field which is the gradient of the harmonic potential 
$\phi_{1}^{\text{I}}(\lb{x})$,
\begin{equation}
\label{eq:phi1}
\phi_{1}^{\text{I}}(\lb{x}) = i \omega\,\frac{ \mu_{0}}{4\pi}\,
\frac{\lbhat{x}\gsb\cdot(\lbhat{n}\gsb\times\lb{m})}{|\lb{x}|+\lb{x}\cdot\lbhat{n}}
\end{equation}
To satisfy the boundary conditions between Regions~I and II, we 
must also add the E-field of the image potential \begin{equation}
\label{eq:phi2}
\phi_{2}^{\text{I}}(\lb{x})\equiv -\phi_{1}^{\text{I}}(2 d 
\lbhat{n}-\lb{x}) \end{equation} in order that 
$\lb{E}^{\text{I}}=\gsb\nabla 
(\phi_{1}^{\text{I}}+\phi_{2}^{\text{I}})$ be normal to the slab 
surface, thus ensuring that the transverse component of $\lb{E}$ 
remains continuous at the boundary.

To sustain the $\lb{E}^{\text{I}}$, a charge density $\rho^{\text{I}} 
= \epsilon \lbhat{n}\gsb\cdot\lb{E}^{\text{I}}$ must exist on the 
surface of the slab, and this surface charge must in turn be created 
by a normal-to-surface current within the slab $\lb{j}^{\text{I}} = 
i\omega \rho^{\text{I}}\lbhat{n}$.  Relative to the intraslab eddy 
currents current $\sigma \lb{E}$ which are our main concern, it is 
easy to show that $\lb{j}^{\text{I}}$ is of order 
$\omega\epsilon/\sigma \sim 10^{-9}$ (see the discussion following 
\eqref{eq:Maxwell}) and thus is negligible.


Putting all these pieces together, we are now ready to specify the 
fields in Region~I in a manifestly finite manner, assuming only that 
the spatial size of the current source is reasonably small compared to 
the distance $d$ separating the source from the slab (like the current 
loop in Fig.~\ref{fig:SkinCoil}).\footnote{Also, we continue to ignore 
the displacement current $i\omega\lb{D}$ in Ampere's law 
$\gsb\nabla\gsb\times\lb{H} = \lb{j}+i\omega\lb{D}$; this is 
equivalent to the near-field assumption $\omega d \ll c$, where $c = 
(\mu_{0}\epsilon_{0})^{-1/2}$ is the speed of light.  This near-field 
approximation is very well satisfied for most practical problems in 
quantum-coherent engineering.}

From the 
localized current source, we compute a vector potential 
$\lb{A}(\lb{x})$ in the usual manner
\begin{equation}
\label{eq:A}
\lb{A}(\lb{x}) = \frac{\mu_{0}}{4\pi}\ 
\int_{V}\!\!d^{3}\lbprime{x}\ 
\frac{\lb{j}(\lbprime{x})}{|\lb{x}-\lbprime{x}|}\sim
\frac{\mu_{0}}{4\pi}\ \frac{\lb{m}\gsb\times\lbhat{x}}{|\lb{x}|^{2}}
\end{equation}
where $\lb{j}(\lbprime{x})$ is the (finite) current density within the 
(finite) source.  From the far-field limit of $\lb{A}(\lb{x})$ we 
determine the dipole moment $\lb{m}$ of the source, and under the 
assumption that the size of the source is small compared to $d$, we 
use $\lb{m}$ to compute the back-reaction vorticity field 
$\psi^{\text{I}}(\lb{x})$ via equations 
(\ref{eq:inverse}--\ref{eq:BesselCoefficientResults}) and the 
back-reaction irrotational field $\phi_{2}^{\text{I}}$ via 
\eqref{eq:phi2}.  With the help of the identity 
$-i\omega\lb{A}(\lb{x})=-i\omega\gsb\nabla\gsb\times(\psi^{\text{s}}(\lb{x}))+\gsb\nabla\phi_{1}^{\text{I}}(\lb{x})$, 
which is exact for a point dipole current source, the net B-field and 
E-field in Region~I can be written as follows
\begin{subequations}
\label{eq:finiteFields}
\begin{align}
\lb{B}(\lb{x}) &=\gsb\nabla\gsb\times\lb{A}(\lb{x}) - 
\gsb\nabla(\lbhat{n}\gsb\cdot\gsb\nabla\psi^{\text{I}}(\lb{x}))\\
\lb{E}(\lb{x}) &=-i\omega\lb{A}(\lb{x}) 
   -i\omega\gsb\nabla\gsb\times(\psi^{\text{I}}(\lb{x})\lbhat{n}) +
   \gsb\nabla\phi_{2}^{\text{I}}(\lb{x})
\end{align}
\end{subequations}
where $\lb{A}(\lb{x})$, $\psi^{\text{I}}(\lb{x})$, and 
$\phi_{2}^{\text{I}}(\lb{x})$ are individually nonsingular.

Physically speaking, $\lb{A}(\lb{x})$ describes the E- and B-fields 
that are generated directly by the source current, 
$\psi^{\text{I}}(\lb{x})$ describes the E- and B-fields generated by 
the induced currents within the slab (treating the source current as a 
dipole), and $\phi_{2}^{\text{I}}(\lb{x})$ describes the E-field 
generated by surface charges on the slab (also treating the source 
current as a dipole).  These expressions were used to generate the 
field geometry in Fig.~\ref{fig:SkinCoil}.

\section{Quantum decoherence}
\label{Appendix: Quantum decoherence}
In this section we present an exactly solvable microscopic model of a 
spin-$\tfrac{1}{2}$ particle magnetically coupled to a heat bath.  Our goal is 
derive the expressions for $T_{1}$, $T_{2}$, and $T_{1\rho}$ quoted in 
\eqref{eq:G1covariant}-\eqref{eq:G1rhocovariant}.  


We follow Ford, Lewis, and O'Connell in modeling the heat bath as a 
collection of independent harmonic oscillators \cite{Ford:88}; the 
reader is referred to this article for a thermodynamic justification 
of this model.  We will adjust the bath parameters so as to reproduce 
the known spectral density of magnetic fluctuations, which we computed 
in the previous section.  The quantum Langevin and optical Bloch 
equations---which yield closed-form expressions for $T_{1}$, $T_{2}$, 
and $T_{1\rho}$---will then be uniquely determined.

\subsection{The thermal reservoir {H}amiltonian}
The Hamiltonian $H$ of the spin/heat bath system is taken to be
\begin{equation}
\label{eq:totalHamiltonian}
H = -\gamma \lb{B}_{0}\gsb\cdot\lb{s} + \sum_{j}\frac{1}{2}\omega_{j}\left(
p_{j}^{2}+(q_{j}-\beta_{j}\lbhat{n}_{j}\gsb\cdot\lb{s})^{2}\right)
\end{equation}
Here $\lb{B}_{0}$ is the polarizing field, $\gamma$ is the 
gyromagnetic ratio of the spin, and $\lb{s}$ is the angular 
momentum operator of the spin, satisfying commutation relations 
$[s_{i},s_{j}]=i\hbar \epsilon_{ijk}s_{k}$.  The heat bath is 
described in terms of oscillators with resonant frequency 
$\omega_{i}$, whose dynamical coodinates $\{p_{i},q_{i}\}$ 
satisfy $[q_{i},p_{j}]=i\hbar \delta_{ij}$, which are coupled to 
the spin with strength $\beta_{j}$.  

The physical nature of the heat bath variables need not be otherwise
specified.  Reader may optionally conceive them as, \emph{e.g.},
thermal excitations of conduction band electrons, magnons in a
ferromagnet, excited states of the vacuum in field theory, or in
general as excitations of whatever heat bath is furnished by a given
quantum technology.


We begin by solving the equation of motion of the heat bath 
variables.  In the Heisenberg picture we have\footnote{We recall 
that the Heisenberg equation of motion for an arbitrary operator 
$O(t)$ is $\sdot{O}(t) = [O,H]/(i\hbar)$.  
The results of this section all follow directly from this 
equation, using only the commutation relations for 
$\{q_{i},p_{i},\lb{s}\}$, plus the fact that commutators in the 
Heisenberg representation are the same as those in the 
Schroedinger representation.}
\begin{equation}
\sddot{q}_{j} + \omega_{j}^{2}q_{j} = \beta_{j}\omega_{j}^{2}\lbhat{n}_{j}\gsb\cdot\lb{s}
\end{equation}
Following Ford \emph{et al.}, we write the formal solution to these equations as
\begin{equation}
\label{eq:FormalSolution}
q_{j}(t) = q_{j}^{\text{h}}(t) + \beta_{j}\lbhat{n}_{j}\gsb\cdot\lb{s}(t)-
\int_{-\infty}^{t}\!\!d\sprime{t}\, 
\cos(\omega_{j}(t-\sprime{t}))\,\beta_{j}\lbhat{n}_{j}\gsb\cdot\lbdot{s}(\sprime t)
\end{equation}
Here $q_{j}^{\text{h}}(t)$ is a homogenous solution to the 
equations of motion (corresponding physically to the evolution 
of the heat bath in the absence of the back-action of the spin).
With the help of this result, the spin operator equation of motion 
$\sdot{\lb{s}} = [\lb{s},H]/(i\hbar)$ can be readily cast into the 
form of the following \emph{quantum Langevin equation}\footnote{As a 
help to students, we note that the quantum Langevin equation can be 
written in several equivalent forms, which arise because the 
commutation relation $[q_{j}(t),\lb{s}(t)]=0$ allows the heat bath 
interaction to be written in any of the following fully equivalent 
forms\[ q_{j}(t) \lb{s}(t) = \lb{s}(t) q_{j}(t) = \frac{1}{2} 
(q_{j}(t) \lb{s}(t) + \lb{s}(t) q_{j}(t))\] Upon substituting 
\eqref{eq:FormalSolution} for $q_{j}(t)$, these various forms yield 
fully equivalent (but superficially quite different) Langevin 
equations.  We have chosen to work with the fully symmetrized Langevin 
equation; none of our main results depend on this choice.}
\begin{align}
\label{eq:QuantumLangevinEquation}
\lbdot{s}(t) & = 
	-\gamma\lb{B}_{0}\gsb\times\lb{s}(t) -   
\\\notag&\quad 
		\frac{1}{2}\gamma\big(\lb{B}(t)\gsb\times\lb{s}(t)-
			\lb{s}(t)\gsb\times\lb{B}(t)\big) +
\\\notag&\quad 		
		\frac{1}{2}\,\left[
		\big(\lb{C}\gsb\cdot\lb{s}(t)\big)\gsb\times\lb{s}(t)
		- \lb{s}(t)\gsb\times\big(\lb{C}\gsb\cdot\lb{s}(t)\big)\right]+
\\\notag&\quad 
		 \!\!\int_{-\infty}^{t}\!\!\!\!d\sprime{t}\,\frac{1}{2}
	\left[
		\big(\lb{G}(t-\sprime t)\gsb\cdot\lbdot{s}(\sprime t)\big)
			\gsb\times\lb{s}(t)-
		\lb{s}(t)\gsb\times
			\big(\lb{G}(t-\sprime t)\gsb\cdot\lbdot{s}(\sprime t)\big)
	\right] 
\end{align}
Here $\lb{B}(t)$ is a fluctuating thermal magnetic field which depends only
on the homogenous heat bath operators $q_{j}^{\text{h}}(t)$
\begin{align}
\label{eq:thermalB}
\lb{B}(t) &= \sum_{j}\gamma^{-1}\omega_{j}\beta_{j}q_{j}^{\text{h}}(t) \lbhat{n}_{j}\\
\intertext{and $\lb{G}(t-\sprime t)$ is a dissipative kernel} 
\label{eq:Gdef}
\lb{G}(t-\sprime t) & =
\begin{cases}
\sum_{j}\omega_{j}\beta_{j}^{2}\,\cos(\omega_{j}(t-\sprime t))\,\lbhat{n}_{j}\gsb\otimes\lbhat{n}_{j}&
t>\sprime t\\
0&t<\sprime t
\end{cases}
\end{align} 
Note that $\lb{G}(t-\sprime t)$ is a pure c-number matrix which 
is independent of the temperature of the heat bath.  It is the 
matrix generalization of what Ford \emph{et al.} call the 
``memory function'' of the heat bath.

The \emph{anisotropy tensor} $\lb{C}$ is a symmetric c-number tensor 
given by
\begin{equation}
	\label{eq:anisotropy}
	\lb{C} = \sum_{j} \omega_{j}\beta_{j}^{2} \,
	\lbhat{n}_{j}\gsb\otimes\lbhat{n}_{j}
\end{equation}
Physically speaking, $\lb{C}$ describes the interaction of the 
particle with image currents in the nearby slab.  This reflects a very 
general and physically realistic property of the independent 
oscillator model: the dynamical equations of a particle are 
\emph{renormalized} by its interaction with the thermal reservoir.  

For the special case of spin-$\tfrac{1}{2}$ particles $\lb{C}$ has no 
dynamical effects, due to an identity satisfied by spin-$\tfrac{1}{2}$ 
operators (and not by higher-spin particles)
\begin{equation}
	s_{i}s_{j} + s_{j}s_{i} = \frac{\hbar^{2}}{2}\,\lbb{I}
\end{equation}
where $\lbb{I}$ is the identity operator.  This identify 
ensures that the $\lb{C}$-dependent terms in 
\eqref{eq:QuantumLangevinEquation} vanish identically; we emphasize 
that this occurs only for spin-$\tfrac{1}{2}$ particles.

So far, we have not used the fact that the heat bath variables 
are in thermal equilibrium; we now use this fact to compute the 
spectral density $\lb{S}_{\lb{B}}$ of $\lb{B}(t)$.  Not 
surprisingly, we will discover that $\lb{S}_{\lb{B}}$ has a 
functional form that is closely related to the dissipative kernel 
\lb{G}.

The calculation is straightforward.  Denoting a thermal ensemble 
average by $\expect{\ldots}_{t}$, we follow Ford \emph{et al.} in 
noting that the correlation of the homogenous heat bath operators 
$\{q_{i}^{\text{h}}\}$ is of the simple form
\begin{equation}
\expect{q_{i}^{\text{h}}(t)q_{j}^{\text{h}}(\sprime t)}_{t} = \frac{1}{2}\,\delta_{ij}\,
\hbar\coth\left(\frac{\hbar\omega_{j}}{2k_{\text{B}}T}\right)\,\cos(\omega_{j}(t-\sprime{t}))
\end{equation}
Combining this with \eqref{eq:thermalB} yields an explicit 
expression for $\lb{S}_{\lb{B}}(\omega)$
\begin{align}
\label{eq:heatBath}
\lb{S}_{\lb{B}}(\omega) &= \int_{-\infty}^{\infty}\!\!d\tau\,e^{-i\omega \tau}\frac{1}{2}
\expect{\lb{B}(t)\gsb\otimes\lb{B}(t+\tau)+\lb{B}(t+\tau)\gsb\otimes\lb{B}(t)}_{t}\\ \notag
&=\sum_{j}\frac{\omega_{j}^{2}\beta_{j}^{2}}{2\gamma^{2}}\,
\lbhat{n}_{j}\gsb\otimes\lbhat{n}_{j}\,
\hbar\coth\left(\frac{\hbar\omega}{2k_{\text{B}}T}\right)\,
\int_{-\infty}^{\infty}\!\!d\sprime t\,e^{-i\omega \tau} 
\cos(\omega_{j}\tau)\\ \notag
 &= \sum_{j}\frac{\pi\omega_{j}\beta_{j}^{2}}{2\gamma^{2}}\,
 \lbhat{n}_{j}\gsb\otimes\lbhat{n}_{j}\,
 \hbar\omega \coth\left(\frac{\hbar\omega}{2k_{\text{B}}T}\right)\,
 \big(\delta(\omega-\omega_{j})+\delta(\omega+\omega_{j})\big)
\end{align}
If we compare this result with the real part of the 
Fourier transform of the dissipative kernel 
\begin{align}
	\label{eq:dissipativeKernel}
\text{Re}\,\lbtilde{G}(\omega) &= \text{Re} 
\int_{-\infty}^{\infty}\!\!d\sprime t\,e^{-i\omega(t-\sprime t)}\lb{G}(t-\sprime t)\\ \notag
&= \frac{\pi}{2}\,\sum_{j}\omega_{j}\beta_{j}^{2}\,
\lbhat{n}_{j}\gsb\otimes\lbhat{n}_{j}\,
 \big(\delta(\omega-\omega_{j})+\delta(\omega+\omega_{j})\big)
\end{align}
it is apparent that
\begin{equation}
\label{eq:FDTheoremForSpins}
\lb{S}_{\lb{B}}(\omega) = \gamma^{-2} \text{Re}\,\lbtilde{G}(\omega)\,
\hbar\omega \coth\left(\frac{\hbar\omega}{2k_{\text{B}}T}\right)
\end{equation}
In terms of the magnetic dissipation matrix $\gsb{\Gamma}(\omega)$ 
computed in the previous section, this takes an even simpler form
\begin{equation}
\label{eq:FDTheoremForSpins2}
\gamma^{2}\gsb{\Gamma}(\omega) = \text{Re}\,\lbtilde{G}(\omega)
\end{equation}
This is the fluctuation-dissipation theorem as it applies to 
particles of arbitrary spin.  For our purposes an alternative 
relation is even more useful: if $\gsb{\Gamma}(\omega)$ is known, 
then the dissipative kernel $\lb{G}(t-\sprime t)$ is given 
explicitly by
\begin{equation}
\label{eq:FDTheoremForSpins3}	
\lb{G}(t-\sprime t) =
\begin{cases}
{\displaystyle \frac{\gamma^{2}}{\pi} \int_{-\infty}^{\infty}\!\!d\omega\,
e^{i\omega(t-\sprime t)}\gsb{\Gamma}(\omega)}, & t>\sprime t;\\[1.5ex]
0,&t<\sprime t;
\end{cases}
\end{equation}
as may be verified by writing $\gsb{\Gamma}(\omega)$ in terms of 
$\lb{S}_{\lb{B}}(\omega)$ using \eqref{eq:GammaNoise}, then 
writing $\lb{S}_{\lb{B}}(\omega)$ in terms of heat bath variables 
using \eqref{eq:heatBath}, then carrying out the integration and 
comparing with \eqref{eq:Gdef}.\footnote{ It may occur to the 
reader that by virtue of \eqref{eq:FDTheoremForSpins2} and 
\eqref{eq:FDTheoremForSpins3}, knowledge of 
$\text{Re}\,\stilde{\lb{G}}(\omega)$ suffices to determine 
$\text{Im}\,\stilde{\lb{G}}(\omega)$.  This insight leads 
directly to a \emph{Kramers-Kroenig} relation:
\begin{align}\notag
	\text{Im}\,\stilde{\lb{G}}(\omega) &= 
	\text{Im}
	\int_{-\infty}^{\infty}\!\!d\tau\,
	e^{-i\omega\tau}\lb{G}(\tau)\\
\notag
	&= 
	\text{Im}
	\int_{0}^{\infty}\!\!d\tau
	\int_{-\infty}^{\infty}\!\!d\sprime\omega\,
	e^{i(\sprime\omega-\omega)\tau}
	\frac{\gamma^{2}}{\pi}\gsb{\Gamma}(\sprime\omega)\\
\notag
	&= \frac{1}{\pi}
	\int_{0}^{\infty}\!\!d\tau
	\int_{-\infty}^{\infty}\!\!d\sprime\omega\,
	\sin\big((\sprime\omega-\omega)\tau\big)\,
	\text{Re}\,\stilde{\lb{G}}(\sprime\omega)\\
\intertext{which is the Kramers-Kroenig relation for particles of 
arbitrary spin interacting with thermal magnetic noise.  Most 
textbooks present such relations in a more compact but less 
obviously finite form, which can be obtained by rewriting the 
above result as} \notag &= \frac{1}{\pi} 
\int_{0}^{\infty}\!\!d\tau\ \frac{d}{d\tau}\!  
\int_{-\infty}^{\infty}\!\!d\sprime\omega\, 
\cos\big((\sprime\omega-\omega)\tau\big)\, 
\frac{\text{Re}\big(\stilde{\lb{G}}(\sprime\omega)-\stilde{\lb{G}}(\omega)\big)}{\omega-\sprime{\omega}}\\
\intertext{where the $\stilde{\lb{G}}(\omega)$ term is introduced 
to regulate the singularity at $\sprime\omega = \omega$, and the 
$\sprime\omega$ integration contour is adjusted in the complex 
plane so that the net contribution of this added term is zero.  
The $\tau$ integration then becomes trivial and the remaining 
$\sprime \omega$ integration is the Kramers-Kroenig relation in 
its traditional form}
\label{eq:KramersKronig}
\text{Im}\,\stilde{\lb{G}}(\omega) &= 
	\frac{1}{\pi} \ P\!\!\int_{-\infty}^{\infty}\!\!d\sprime\omega\,
	\frac{\text{Re}\,\stilde{\lb{G}}(\sprime\omega)}{\omega-\sprime{\omega}}\,
\end{align} where $P$ is a principal value.}

\subsection{The master equation}
We now direct our attention to solving the quantum Langevin equation 
\eqref{eq:QuantumLangevinEquation}.  As it stands, this equation is 
too complicated to solve directly.\footnote{Students (and engineers) 
may wish to reflect that the quantum Langevin 
equation~\eqref{eq:QuantumLangevinEquation} is easy to solve in 
principle---it can be numerically integrated quite readily (the 
integration is quite straightforward to program in languages like 
\emph{Mathematica}; this is a good exercise for a student).  
Heisenberg operators are be represented numerically by matrices of 
complex numbers, with operator addition and multiplication achieved 
via ordinary matrix addition and multiplication.  As time goes on, 
the quantum Langevin equation increaingly entangles the spin operator $\lb{s}(t)$ 
with the thermal reservoir basis states---thus leading, as expected, to 
fluctuation, dissipation, and entanglement of the spin operators.

The only practical difficulty with this program is that a reservoir of 
$n$ particles, each having $m$ quantum states, requires $m^{n}$ 
quantum basis states, with each Heisenberg operator stored as a 
time-dependent $m^{n}\times m^{n}$ Hermitian matrix.  For $n$ as small 
as two and $m$ as small as twenty---too small to describe a realistic 
thermal reservoir---the storage and multiplication of these 
exponentially large matrices is enough to overwhelm \emph{any} 
classical computer.  Quantum computers were invented in part to 
overcome this fundamental \emph{quantum simulation problem}.}
In particular, $B(t)$ is an operator-valued function of 
heat bath variables whose detailed dynamical behavior---describing 
every minute fluctuation within the thermal reservoir---we neither 
know nor wish to know.  To make progress, we will average over a 
thermodynamic ensemble of heat bath variables, and study only a 
coarse-grained time derivative of $\lb{s}(t)$.  Simplified equations 
obtained by this general strategy are called \emph{master 
equations}---our goal is to derive a master equation from the quantum 
Langevin equation.

We begin by defining a rotation matrix $\lb{R}(t)$, satisfying 
$\lbdot{R}(t) = \gamma\lb{B}_{0}\gsb\times\lb{R}(t)$, in terms of which 
we define rotating-frame quantities
\begin{align}
\notag
\lb{s}_{\text{rot}}(t) &\equiv \lb{R}(t)\gsb\cdot\lb{s}(t)\\
\notag
\lb{B}_{\text{rot}}(t) &\equiv \lb{R}(t)\gsb\cdot\lb{B}(t)\\
\notag
\lb{G}_{\text{rot}}(t-\sprime t) &\equiv \lb{R}(t)\gsb\cdot\lb{G}(t-\sprime t)\gsb\cdot\lb{R}^{\dagger}(t)
\end{align}
It follows that $\lbdot{s}(t) = 
\lb{R}^{\dagger}(t)\gsb\cdot(\lbdot{s}_{\text{rot}}(t)-\gamma\lb{B}_{0}\gsb\times\lb{s}_{\text{rot}}(t))$. 

Neglecting $\lcal{O}(\lbdot{s}_{\text{rot}}^{2})$ terms---this 
is the rotating-frame approximation---and specializing  to
spin-$\tfrac{1}{2}$ particles so that the anisotropy tensor
$\lb{C}$ does not enter, the quantum Langevin equation 
\eqref{eq:QuantumLangevinEquation} becomes
\begin{multline}
\label{eq:RotatingQuantumLangevinEquation}
\lbdot{s}_{\text{rot}}(t) = -
\frac{1}{2}\gamma\big(\lb{B}_{\text{rot}}(t)\gsb\times\lb{s}_{\text{rot}}(t)-
\lb{s}_{\text{rot}}(t)\gsb\times\lb{B}_{\text{rot}}(t)\big) + \\
\frac{1}{2}\int_{-\infty}^{t}\!\!d\sprime{t}\,
\Big(
\big(\lb{G}_{\text{rot}}(t-\sprime t)\gsb\cdot(\gamma\lb{B}_{0}
\gsb\times\lb{s}_{\text{rot}}(\sprime t))\big)\gsb\times\lb{s}_{\text{rot}}(\sprime t)-\\
\lb{s}_{\text{rot}}(\sprime t)\gsb\times\big(\lb{G}_{\text{rot}}(t-\sprime t)
\gsb\cdot(\gamma\lb{B}_{0}\gsb\times\lb{s}_{\text{rot}}(\sprime t))\big)
\Big)
\end{multline}
Recalling that $\lb{G}_{\text{rot}}(\tau)$ is the 
temperature-independent ``memory function'' of the heat bath, we 
now assume it decorrelates rapidly compared to the rate at 
which $\lb{s}_{\text{rot}}(t)$ varies.  We can then 
ignore the time-dependence of $\lb{s}_{\text{rot}}(t)$ 
inside of integrals, and make the following substitution
\begin{equation}
\label{eq:substituteA}
	\int_{-\infty}^{t}\!\!d\sprime t\ \lb{G}_{\text{rot}}(t-\sprime t) =
	\gamma^{2} \gsb\Gamma_{\text{rot}}(0)
\end{equation}
Here the rotating-frame magnetic dissipation matrix 
$\gsb\Gamma_{\text{rot}}(\omega)$ is related to the 
laboratory-frame matrix $\gsb\Gamma(\omega)$ via
	\begin{multline}
\qquad \gsb\Gamma_{\text{rot}}(\omega) =
	\big[\lbhat{b}\gsb\otimes\lbhat{b}\big]\ 
	\text{tr}\big[\lbhat{b}\gsb\otimes\lbhat{b}
	\gsb\cdot\gsb\Gamma(\omega)\big]+\\
	\qquad\qquad\big[\lb{I}- \lbhat{b}\gsb\otimes\lbhat{b}\big]\ \frac{1}{2}	
	\,\text{tr}\big[(\lb{I}- 
\lbhat{b}\gsb\otimes\lbhat{b})
\gsb\cdot\gsb\Gamma(\omega_{0})\big]\qquad
\end{multline}
with $\gamma \lb{B}_{0} \equiv \omega_{0}\lbhat{b}$.  Similarly, 
$\lb{S}_{\lb{B}_{\text{rot}}}(\omega)$ is related to 
$\lb{S}_{\lb{B}}(\omega)$ via \eqref{eq:RotatingSpectralDensity}.  
Then, with the help of some straightforward algebraic 
manipulations,\footnote{The derivation involves
integrating \eqref{eq:RotatingQuantumLangevinEquation} to first 
order in $\lb{G}_{\text{rot}}$ and second order in 
$\lb{B}_{\text{rot}}$, then substituting \eqref{eq:substituteA}.  
The rotating-frame commutation relations are then invoked
to simplify the cross-terms:
\[
[(\lb{s}_{\text{rot}}(t))_{i},(\lb{s}_{\text{rot}}(t))_{j}] = i\hbar 
\epsilon_{ijk}(\lb{s}_{\text{rot}}(t))_{k}, \] These are valid because 
spin commutators are form-invariant under transformation to 
the Heisenberg picture and to the rotating frame.  
Further specializing to spin-$\tfrac{1}{2}$ implies
\[
\left(\lb{s}_{\text{rot}}(t)\right)_{x}^{2} =
\left(\lb{s}_{\text{rot}}(t)\right)_{y}^{2} =
\left(\lb{s}_{\text{rot}}(t)\right)_{z}^{2} 
= \left(\frac{\hbar}{2}\right)^{2}\,\lbb{I}
\]
where $\lbb{I}$ is the identity operator; these identities also 
are form-invariant.  Finally, when we average over a thermal 
ensemble, we have 
\[
\int_{-\infty}^{\sprime t}\!\!d t'\,\frac{1}{2}\,
\expect{\lb{B}_{\text{rot}}(t)\gsb\otimes\lb{B}_{\text{rot}}(t')+t\leftrightarrow 
t'} = \lb{S}_{\lb{B}_{\text{rot}}}(0)
\]
as well as (trivially) $\expect{\lbb{I}}=1$.} the Langevin equation 
for spin-$\tfrac{1}{2}$ particles can be cast into the form of a master equation, 
which turns out to be the following \emph{Bloch equation}
\begin{align}
	\label{eq:FullBlochEquations}
	\frac{d}{d\,t}\,\expect{\lb{s}_{\text{rot}}(t)} =\ & -\frac{1}{T_{1}}\ 
	\lbhat{b}\gsb\otimes\lbhat{b}\gsb\cdot\left(
	\expect{\lb{s}_{\text{rot}}(t)}-\expect{\lb{s}_{0}}\right)\\
\notag
	&-\frac{1}{T_{2}}\ 
	(\lb{I}-\lbhat{b}\gsb\otimes\lbhat{b})\gsb\cdot\expect{\lb{s}_{\text{rot}}(t)}
\end{align}
Here $T_{1}$ and $T_{2}$ are as given in \eqref{eq:relaxations}, 
and we have wrapped the spin operator in an ensemble average 
$\expect{\ldots}$ in order to express the Bloch equations in 
terms of the (c-number) vector spin polarization 
$\expect{\lb{s}_{\text{rot}}}$, as is traditional.  The 
equilibrium spin polarization $\expect{\lb{s}_{0}}$ is found to be
\begin{equation}
\expect{\lb{s}_{0}} = \frac{\hbar}{2}\,\tanh\left(\frac{\hbar 
\gamma B_{0}}{2 k_{\text{B}}T}\right)\lbhat{b}
\end{equation}
which agrees with the well-known thermodynamic expression for 
spin-$\tfrac{1}{2}$ polarization; this provides a consistency check of 
the Langevin formalism.

The calculation of the spin-locked relaxation time $T_{1\rho}$ is 
similar, differing only in that transformation to a doubly 
rotating frame is required.  The result is as given in 
\eqref{eq:relaxations}; details of the derivation will not be 
given.

We close by noting that our spin relaxation rates differ from the 
oscillator relaxation rates of Ford \emph{et al.} \cite{Ford:88} in 
one physically important respect: $T_{1}$, $T_{2}$, and $T_{1\rho}$ 
are strongly temperature-dependent, while Ford \emph{et al.} predict 
an oscillator quality $Q$ that is independent of temperature.  
Formally, the reason for this difference can be traced to the Langevin 
equation \eqref{eq:RotatingQuantumLangevinEquation}, in which the 
dissipative kernel $\lb{G}_{\text{rot}}$ is temperature-independent 
(this is true for both spins and oscillators).  However, when this 
Langevin equation is integrated to second order to obtain the Bloch 
equations, the temperature-dependent fluctuations $\lb{B}(t)$ 
\emph{also} contribute to spin relaxation via the commutation relation 
$[s_{i},s_{j}] = i\hbar \epsilon_{ijk}s_{k}$.  In contrast, the 
oscillator operators $p$ and $q$ have a pure c-number commutator 
$[q,p]=i\hbar$, and in consequence the fluctuating force exerted by 
the reservoir does \emph{not} directly contribute to oscillator 
relaxation.

The difference between spin relaxation and oscillator relaxation can
also be understood physically.  If we imagine a classical spin that is
subject to a randomly fluctuating magnetic field, and no other
physical influence, the average spin orientation $\expect{\lb{s}}$
necessarily will relax toward $\expect{\lb{s}}=0$.  So to the extent
that a higher-temperature reservoir creates stronger magnetic
fluctuations, spins will relax more rapidly at higher temperature.  In
contrast, a randomly fluctuating force does not change the average
trajectory of an oscillator at all---provided the average force is
zero---because oscillator equations of motion are linear, unlike spin
equations.  So to the extent that a thermal reservoir has a
temperature-independent dissipative kernel---like the independent
oscillator model---the oscillator quality $Q$ will be independent of
temperature.

\section{Fluctuation-dissipation-entanglement theorems}
\label{Appendix: A fluctuation-dissipation-entanglement theorem}

Now we will prove fluctuation-dissipation-entanglement theorems
for both spin-$\tfrac{1}{2}$ particles and harmonic oscillators in 
contact with a thermal reservoir.

\subsection{A spin-$\tfrac{1}{2}$ fluctuation-dissipation-entanglement 
theorem}

Section~\ref{Section: A fluctuation-dissipation-entanglement theorem} 
has already provided the definitions we need to prove the 
fluctuation-dissipation-entanglement theorem, and 
Appendix~\ref{Appendix: Quantum decoherence} has carried 
through many of the needed calculations.  We need only organize our 
reasoning as follows.

Specializing to spin-$\tfrac{1}{2}$ particles, we write the total 
Hamiltonian~\eqref{eq:totalHamiltonian} in the usual form $H = H_{0} + 
V$, with the perturbing Hamiltonian $V$ given by
\begin{equation}
	\label{eq:V}
V = -\sum_{j}\omega_{j} \beta_{j}q_{j}\,\lbhat{n}_{j}\gsb\cdot\lb{s}
\end{equation}
The exact ground state $\ket{\Psi_{0}}$ of $H$ can be calculated 
order-by-order in $V$ as
\[ 
    \ket{\Psi_{0}} = \ket{\Psi_{0}^{(0)}} + 
    \ket{\Psi_{0}^{(1)}} + \ket{\Psi_{0}^{(2)}} + \ldots
\] 
in the notation of Landau and Lifschitz \cite{Landau:58}.  The formal 
expression for $\ket{\Psi_{0}^{(2)}}$ is quite lengthy, but 
fortunately we have $\lcal{E}\ket{\Psi_{0}^{(0)}}=0$ by construction 
\eqref{eq:defOfE}, so that cross terms like 
$\braket{\Psi^{(2)}_{0}}{\lcal{E}}{\Psi_{0}^{(0)}}$ vanish.  In 
consequence, the leading contribution to the entanglement comes 
entirely from $\braket{\Psi_{0}^{(1)}}{\lcal{E}}{\Psi_{0}^{(1)}}$,
for which we have the well-known expression
\begin{equation}
	\ket{\Psi_{0}^{(1)}} = \sum_{m\ne 0} 
	\frac{\braket{\psi^{(0)}_{m}}{V}{\psi^{(0)}_{0}}}{E_{0}-E_{m}}\,
	\ket{\psi^{(0)}_{m}}
\end{equation}
The fluctuation-dissipation-entanglement theorem \eqref{eq:fde
theorem} asserted in Section~\ref{Section: A
fluctuation-dissipation-entanglement theorem} can now be readily
derived
\begin{subequations}
\begin{align}
	\label{eq:StartFDETheorem}
	\qquad& \hspace{-2em}E(\lbhat{p},\omega_{0}) \equiv 
	\braket{\Psi_{0}}{\lcal{E}}{\Psi_{0}} 
	\simeq \braket{\Psi_{0}^{(1)}}{\lcal{E}}{\Psi_{0}^{(1)}}
	& & 
\\
	&=\sum_{j} 
	\frac{\hbar\omega_{j}^{2}\beta_{j}^{2}}
	{8(\omega_{0}+\omega_{j})^{2}}\,
	\text{tr}\big[(\lb{I}-\lbhat{p}\gsb\otimes\lbhat{p})\gsb\cdot
	(\lbhat{n}_{j}\gsb\otimes\lbhat{n}_{j})\big] 
	& & \text{by \eqref{eq:V}}
\\
    \label{eq:DummyIntegration}
	&=\frac{\hbar}{4\pi}\!\int_{0}^{\infty}\!\!\!d\omega\, 
	\frac{\omega }{(\omega_{0}+\omega)^{2}}\,
	\begin{array}[t]{l}\displaystyle
	\biggl(
     \sum_{j}\frac{\pi\omega_{j}\beta_{j}^{2}}{2}\,
	 \delta(\omega-\omega_{j})\\
	\hspace{-2.5em}\text{tr}\big[(\lb{I}-\lbhat{p}\gsb\otimes\lbhat{p})\gsb\cdot
	(\lbhat{n}_{j}\gsb\otimes\lbhat{n}_{j})\big]\biggr)
	\end{array}
	& &
\\
\label{eq:analyticProperties}
	&=\frac{\hbar}{4 \pi}\ 
    \text{\large $\lfrak{G}$}_{2}\big\{\omega\,\text{tr}\big[
	(\lb{I}-\lbhat{p}\gsb\otimes\lbhat{p})
	\gsb\cdot\text{Re}(\lbtilde{G}(\omega))\big];\omega_{0}\big\}
	& & \text{by \eqref{eq:dissipativeKernel}}
\\
	&=\frac{\hbar\gamma^{2}}{4 \pi}\ 
	\text{\large $\lfrak{G}$}_{2}\big\{\omega\,\text{tr}\big[
	(\lb{I}-\lbhat{p}\gsb\otimes\lbhat{p})
	\gsb\cdot\gsb{\Gamma}(\omega)\big];\omega_{0}\big\}
    & & \text{by \eqref{eq:FDTheoremForSpins2}}
\end{align}
\end{subequations}
where \eqref{eq:DummyIntegration} was obtained by introducing $\omega$ 
as a dummy variable of integration over $\delta(\omega-\omega_{j})$.

\subsection{An oscillator fluctuation-dissipation-entanglement
theorem} A similar fluctuation-dissipation-entanglement theorem exists
for harmonic oscillators.  We specify the Hamiltonian $H$ of the
oscillator/thermal reservoir as
\begin{equation}
\label{eq:totalOscillatorHamiltonian}
H = \frac{1}{2}\omega_{0}\left(p^{2} + 
q^{2}\right) + 
\sum_{j}\frac{1}{2}\omega_{j}\beta_{j}\left(
p_{j}^{2}+(q_{j}-q)^{2}\right)
\end{equation}
Here $q$ and $p$ are generalized oscillator coordinates 
satisfying $[q,p] = i\hbar$, whose physical nature is not otherwise 
specified.  As shown by Ford, Lewis, and O'Connell \cite{Ford:88}, the 
Hamiltonian \eqref{eq:totalOscillatorHamiltonian} implies the 
following quantum Langevin equation in the Heisenberg picture
\begin{equation}
	\label{eq:oscillatorDynamics}
	\sddot{q}(t) + \omega_{0}^{2}q(t) + \omega_{0}\!\!
	\int_{-\infty}^{t}\!\!\!\!d\sprime{t}\,
	\mu(t-\sprime{t})\sdot{q}(\sprime{t}) = 
	f(t)
\end{equation}
where $f(t)$ is a fluctuating thermal force, and $\mu(\tau)$ is a 
dissipative ``memory function'' given by
\begin{equation}
	\mu(\tau) = 
	\begin{cases}
		\sum_{j} \omega_{j}\beta_{j}^{2} \cos(\omega_{j}\tau)&\text{for}\ \tau\ge0;\\
		0 &\text{for}\ \tau<0.
	\end{cases}
\end{equation}
The Fourier transform of $\mu(\tau)$
\begin{equation}
	\stilde{\mu}(\omega)\equiv\int_{-\infty}^{\infty}\!\!\!d\tau\,
		e^{i\omega\tau}\mu(\tau)
\end{equation}
has a real part given explicitly by
\begin{equation}
	\label{eq:mutilde}
	\text{Re}[\stilde{\mu}(\omega)] = \frac{\pi}{2}\,\sum_{j} 	
	\omega_{j}\beta_{j}^{2}
	\big(\delta(\omega-\omega_{j})+\delta(\omega+\omega_{j})\big)
\end{equation}
which is related to the spectral density $S_{f}(\omega)$ of the thermal 
force $f(t)$ by
\begin{equation}
	\label{eq:SfFor Oscillators}
	S_{f}(\omega) = \omega_{0}^{2}\,\text{Re}[\stilde{\mu}(\omega)]\,
	\hbar\omega\,\coth\left(\frac{\hbar\omega}{2 k_{\text{B}}T}\right)
\end{equation}
This is, of course, the celebrated fluctuation-dissipation theorem for 
harmonic oscillators. 

Now we have all the pieces we need to calculate the zero-temperature 
entanglement from the dissipative kernel $\stilde{\mu}(\omega)$.  By 
reasoning exactly analogous to the spin-$\tfrac{1}{2}$ case 
(\ref{eq:StartFDETheorem}--\ref{eq:analyticProperties}) we find
\begin{subequations}
\label{eq:harmonicFDE}
\begin{align}
	\braket{\Psi_{0}}{\lcal{E}}{\Psi_{0}} 
	& \simeq \braket{\Psi_{0}^{(1)}}{\lcal{E}}{\Psi_{0}^{(1)}}
	& & 
\\
	&=\sum_{j} 
	\frac{\omega_{j}^{2}\beta_{j}^{2}}
	{4(\omega_{0}+\omega_{j})^{2}}
	& &
\\
 	&=\int_{0}^{\infty}\!\!\!d\omega\, 
	\frac{\omega}
	{(\omega_{0}+\omega)^{2}}\,
	\text{Re}[\stilde{\mu}(\omega)]
	& &
\\
	&=\frac{1}{2\pi}\ 
    \text{\large $\lfrak{G}$}_{2}\big\{
	\omega\text{Re}[\stilde{\mu}(\omega)];\omega_{0}\big\}
	& & 
\end{align}
\end{subequations}
This is the general form of the fluctuation-dissipation-entanglement 
theorem for harmonic oscillators.

As in Section~\ref{Section: Approximate expression for the 
entanglement} for spin-$\tfrac{1}{2}$ particles, oscillator 
entanglement can be evaluated approximately.  We stipulate that 
$\stilde{\mu}(\omega)\sim 1/Q$ for $\omega \in (0,\omega_{c})$, with 
$Q$ the oscillator quality and $\omega_{\text{c}}$ the thermal 
reservoir's cutoff frequency $\omega_{\text{c}}\gg\omega_{0}$.  Then 
we find to leading order in $\omega_{0}/\omega_{\text{c}}$
\begin{equation}
	\label{eq:OhmicOscillator}
	\braket{\Psi_{0}}{\lcal{E}}{\Psi_{0}} \simeq
	\frac{\ln(\omega_{\text{c}}/\omega_{0})}{2\pi Q}\,
\end{equation}
This approximate result is not new.  Li, Ford, and O'Connell
\cite[eq.~14]{Ford:95} have calculated a zero-temperature energy shift
that directly implies \eqref{eq:OhmicOscillator}, in the context of
their reply to a critique by Senitzky \cite{Senitzky:95} of a
still-earlier analysis of energy balance in dissipative systems
\cite{Ford:93}.  However, they did not interpret the energy shift as
an entanglement relation, and the generality and rigor of the link
between dissipation and entanglement via \Stieltjies transforms was
not pointed out.

\subsection{Oscillator renormalization}
For spin-$\tfrac{1}{2}$ particles, we have seen that the thermal 
reservoir interactions \emph{renormalize} the dynamical equations of 
the field operators, via the anisotropy tensor $\lb{C}$ 
(\ref{eq:QuantumLangevinEquation},\ref{eq:anisotropy}).  We will now 
show that a similar renormalization occurs for oscillators.

To begin, we remark that in quantum-coherent engineering,
renormalization is more than an abstract concept.  At least in the
context of magnetic resonance force microscopy, we will see that
renormalization effects are large---they play a key role in
experimental protocols---and are of considerable practical
consequence.

Next, we note that oscillator renormalization is much easier to 
analyze if we recognize at the outset that the operators $q$ and $p$ 
and the frequency $\omega_{0}$ which appear in the independent 
oscillator Hamiltonian \eqref{eq:totalOscillatorHamiltonian} are 
\emph{already} renormalized.  We know this from the quantum Langevin 
equation \eqref{eq:oscillatorDynamics}, in which $\omega_{0}$ appears 
as the resonant frequency of the oscillator \emph{after} it has been 
renormalized by interactions with the thermal reservoir.  Our task, 
therefore, is to deduce an unrenormalized ``bare'' Hamiltonian from 
the renormalized ``dressed'' Hamiltonian 
\eqref{eq:totalOscillatorHamiltonian}.  A great virtue of the 
independent oscillator model is that this calculation can be carried 
through quite easily.

We define an unrenormalized frequency $\sprime{\omega_{0}}$ by
\begin{subequations}
\begin{align}
	\label{eq:unrenormalizedOmega}
	\sprime{\omega_{0}}  &\equiv 
	\big[\omega_{0}\,
	\big(\omega_{0}+{\textstyle \sum_{j}}\,\omega_{j}\beta_{j}^{2}\big)
	\big]^{1/2}
	\\
\intertext{and unrenormalized operators $\sprime{q}$ and $\sprime{p}$ by}
	\sprime{q} &\equiv q\,(\sprime{\omega_{0}}/\omega_{0})^{1/2}\\
	\sprime{p} &\equiv p\,(\omega_{0}/\sprime{\omega_{0}})^{1/2}\\
\intertext{and unrenormalized reservoir couplings $\sprime{\beta_{j}}$ by} 
	\label{eq:unrenormalizedBeta}
    \sprime{\beta_{j}} &\equiv 
    \beta_{j}\,(\omega_{0}/\sprime{\omega_{0}})^{1/2}
\end{align}
\end{subequations}
By construction, the unrenormalized operators $\sprime{q}$ and 
$\sprime{p}$ satisfy the canonical commutation relation 
$[\sprime{q},\sprime{p}]=i\hbar$.  Furthermore, the unrenormalized 
parameters $\sprime{\omega_{0}}$ and $\{\sprime{\beta_{j}}\}$ are such 
that the Hamiltonian---when written in terms of unrenormalized 
quantities---takes the desired ``bare'' form 
$H=H_{0}+V$, with $H_{0}$ 
and $V$ given by
\begin{subequations}
\begin{align}
\label{eq:renormalizedOscillatorHamiltonian}
H_{0} & = \frac{1}{2} \sprime{\omega_{0}}\big(\sprime{p}^{2} + 
\sprime{q}^{2}\big) + 
\sum_{j}\frac{1}{2}\omega_{j} \big(p_{j}^{2}+q_{j}^{2}\big)\\
\label{eq:defOfV}
V & = -\sum_{j}\omega_{j}q_{j}\sprime{\beta_{j}}\sprime{q}
\end{align}
\end{subequations}
The virtue of the ``bare'' Hamiltonian $H_{0}$ is that it clearly shows 
what happens when the oscillator-reservoir couplings 
$\{\sprime{\beta_{j}}\}$ are turned off: the resulting ``bare'' 
oscillator dynamics are described in terms of unrenormalized canonical 
operators $\sprime{q}$ and $\sprime{p}$ and resonant frequency 
$\sprime{\omega_{0}}$.

The physical meaning of---and the necessity for---renormalization is 
made clear by the following real-world example.  In magnetic resonance 
force microscopy (MRFM) experiments, the cantilever resonant frequency 
is routinely monitored as the tip of the cantilever is brought near to 
a sample.  As the tip nears the sample, the Van der Waals potential 
(equivalent to the Casimir effect) increasingly acts to reduce the 
cantilever's spring constant, which renormalizes the resonant 
frequency $\omega_{0}$ to a lower value than the ``bare'' value 
$\sprime{\omega_{0}}$, consistent with \eqref{eq:unrenormalizedOmega}.  
Renormalization effects in MRFM are so strong that is commonplace for 
the renormalized frequency to pass through zero and become imaginary, 
in which case the cantilever becomes unstable and the tip ``snaps in'' 
to the sample.  Observation of the renormalized frequency provides a 
vital MRFM technique for monitoring the tip-sample separation without 
ever touching the sample.  In our experiments, we frequently operate
at frequency shifts in excess of 100 Hertz, so that renormalization 
is by no means a small effect.

Concomitantly, the cantilever's quantum ground state is also
renormalized: far from the sample the ground state satisfies
$(\sprime{q}+i\sprime{p})\ket{\sprime{\psi_{0}}}=0$, while near to the
sample the ground state satisfies $(q+ip)\ket{\psi_{0}} = 0$. 
Physically speaking, the renormalized ground state $\ket{\psi_{0}}$
and the unrenormalized ground state $\ket{\sprime{\psi_{0}}}$ are
squeezed relative to one another.  Although deliberately squeezing
cantilever states is not yet routine practice, it is not too
far-fetched to imagine that someday it may be.  

We close this section by remarking that---as is typically the case in
renormalization theory---the inverse problem of computing the
renormalized parameters from the bare parameters is not solvable in
closed form.\footnote{To see this, students should try to
invert (\ref{eq:unrenormalizedOmega}--\ref{eq:unrenormalizedBeta})
to express the renormalized variables $\{q,p,\omega_{0},\beta_{j}\}$
as closed-form functions of the bare variables
$\{\sprime{q},\sprime{p},\sprime{\omega_{0}},\sprime{\beta_{j}}\}$.}
Instead, the renormalized parameters must be calculated
perturbatively, order by order in $\{\sprime{\beta_{j}}\}$, and
convergence is not guaranteed.  In fact, perturbative renormalization
is \emph{guaranteed} to fail when the bare Hamiltonian has a spectrum
that is unbounded from below, because we are trying to convert the
bare Hamiltonian into a manifestly positive-definite renormalized
form.  Such divergences are perfectly physical: they are realized
experimentally whenever a cantilever tip snaps in.  Thus, all aspects
of renormalization theory---even perturbative divergences---acquire
practical significance in quantum-coherent engineering.

\subsection{Renormalization-dissipation relations}
Now we show that renormalization parameters can be calculated from the same 
dissipative kernels that control fluctuation, dissipation, and 
entanglement.  In our nomenclature, these relations do not qualify as 
``theorems'' because the relationship is not invertible---the form of 
the dissipative kernel cannot be inferred from measurements of 
renormalization.

For oscillators, renormalization is entirely specified by the ratio of 
the unrenormalized frequency $\sprime{\omega_{0}}$ to the renormalized 
frequency $\omega_{0}$, per 
(\ref{eq:unrenormalizedOmega}--\ref{eq:unrenormalizedBeta}).  From 
\eqref{eq:mutilde} and \eqref{eq:unrenormalizedOmega} we express the 
frequency shift in terms of the dissipative kernel 
$\text{Re}[\stilde{\mu}(\omega)]$
\begin{align}
	\label{eq:spinRD}
	\frac{\sprime{\omega_{0}}^{2}}{\omega_{0}^{2}} & = 
		1+\sum_{j}\omega_{j}\beta_{j}^{2} \\\notag
		&= 1+\frac{2}{\pi}\!\int_{0}^{\infty}\!\!d\omega\,
		\text{Re}[\stilde{\mu}(\omega)]
\end{align}
For particles of arbitrary spin, renormalization is entirely specified 
by the anisotropy tensor $\lb{C}$, per the quantum Langevin equation 
\eqref{eq:QuantumLangevinEquation}.  From \eqref{eq:anisotropy} amd 
\eqref{eq:dissipativeKernel} we find the anisotropy tensor in terms of 
the dissipative kernel
\begin{align}
	\label{eq:harmonicRD}
	\lb{C} & = \sum_{j} \omega_{j}\beta_{j}^{2} \,
	\lbhat{n}_{j}\gsb\otimes\lbhat{n}_{j}\\\notag
	&=\frac{2}{\pi}\!\int_{0}^{\infty}\!\!d\omega\,
		\text{Re}[\lbtilde{G}(\omega)]
\end{align}
We see that renormalization effects are temperature-independent; this 
is consistent with a careful reading of the independent-oscillator 
literature \cite{Ford:88b}.